\documentclass[12pt]{article}
\usepackage[pdftex]{graphicx}
\usepackage{amssymb}
\usepackage{amsmath}

\newcommand{\rf}[1]{(\ref{#1})}
\newcommand{\beq}{\begin{equation}}
\newcommand{\eeq}{\end{equation}}
\newcommand{\bea}{\begin{eqnarray}}
\newcommand{\eea}{\end{eqnarray}}

\newcommand{\e}{\mbox{e}}

\newcommand{\lam}{\lambda}

\newcommand{\Del}{\Delta}

\newcommand{\kp}{\kappa}
\newcommand{\vxi}{\vec\xi}

\newcommand{\oh}{\frac{1}{2}}

\newcommand{\ra}{\rangle}
\newcommand{\la}{\langle}
\newcommand{\prt}{\partial}

\newcommand{\no}{\nonumber}
\newcommand{\nn}{\no\\}

\newcommand{\Trf}{{\mathbb T}}
\newcommand{\bbT}{\mathbb T}
\newcommand{\K}{\mathbf{K}}
\newcommand{\W}{\mathbf{W}}
\newcommand{\M}{\mathbf{M}}
\newcommand{\boldS}{\mathbf{S}}

\newcommand{\omitt}[1]{ () }
\newcommand{\half}{\frac{1}{2}}
\newcommand{\twothirds}{\frac{2}{3}}
\newcommand{\third}{\frac{1}{3}}
\newcommand{\threehalves}{\frac{3}{2}}
\newcommand{\sixth}{\frac{1}{6}}

\begin{document}

\begin{center}
\vspace{24pt}
{ \large \bf A restricted dimer model on a 2-dimensional random causal triangulation}

\vspace{30pt}

{\sl J. Ambj\o rn}$\,^{a,d}$,
{\sl B. Durhuus}$\,^{b}$,
and {\sl J.F. Wheater}$\,^{c}$

\vspace{48pt}
{\footnotesize

$^a$~The Niels Bohr Institute, Copenhagen University\\
Blegdamsvej 17, DK-2100 Copenhagen \O , Denmark.\\
email: ambjorn@nbi.dk\\

\vspace{10pt}

$^b$~Department of Mathematical Sciences, Copenhagen University,\\
Universitetsparken 5, DK-2100 Copenhagen \O , Denmark.\\
email: durhuus@math.ku.dk\\

\vspace{10pt}

$^c$~Rudolf Peierls Centre for Theoretical Physics, Oxford University, \\
1 Keble Road, Oxford OX1 3NP, UK.\\
email: j.wheater@physics.ox.ac.uk

\vspace{10pt}

$^d$~IMAPP, Radboud University,\\ 
Heyendaalseweg 135,
6525 AJ, Nijmegen, The Netherlands

}
\vspace{48pt}
\end{center}

%\addtolength{\baselineskip}{0.20\baselineskip}
%\vspace{2cm}

\begin{center}
{\bf Abstract}
\end{center}

We introduce a restricted hard dimer model on a random causal 
triangulation that is exactly solvable and generalizes a model recently proposed by Atkin and Zohren \cite{atkin-zohren}. We show 
that the latter model exhibits unusual behaviour at its 
multicritical point; in particular, its Hausdorff dimension equals $3$ 
and not $3/2$ as would be expected from general scaling arguments.  
When viewed as a special case of the generalized model 
introduced here we show that this behaviour is not generic and therefore 
is not likely to represent the true behaviour of the full 
dimer model on a random causal triangulation.

\noindent 
\vspace{12pt}
\noindent \\
%{\bf Draft v8.2: \today }\\
\medskip
\vspace{12pt}

\newpage

\section{Introduction}\label{introduction}

The study of statistical theories of fluctuating geometries
is important for a number of reasons. Regularized via appropriate lattices, so-called
Dynamical Triangulations (DT)\footnote{There are
other ways to provide lattice regularizations of  
bosonic string theory, e.g. using hypercubic lattices \cite{dfj}.}), 
they may serve as rigorous definitions of the path 
integral of bosonic string theories \cite{DT},
or quantum gravity \cite{higherQG}. The DT formalism has
been immensely successful in the study of two-dimensional 
quantum gravity coupled to conformal field theories with 
central change $c \leq 1$, also known as non-critical string 
theory or Liouville quantum gravity, but it has been less 
successful serving as a regularization of a putative higher dimensional 
quantum gravity theory \cite{firstorder}. 
In an attempt to improve the situation 
a modified lattice regularization, called the Causal Dynamical Triangulation (CDT),
was proposed in which a foliation structure is imposed on the 
lattices representing space-time (which we here will assume 
has Eucliean signature) \cite{al,ajl}). Such a foliation
structure is also imposed in the so-called Ho\v{r}ava-Lifshitz gravity
theory \cite{horava}. Some interesting results 
related to higher dimensional quantum gravity have been obtained 
using the CDT regularization (see \cite{physrep} for a review). 
Here we will discuss  the two-dimensional CDT theory, which 
in principle should be simpler than the corresponding %successful 
two-dimensional DT theory. Indeed, the scaling limit of CDT not coupled to 
matter is  simple \cite{al,dgk}, and it can be shown to correspond
to two-dimensional Ho\v{r}ava-Lifshitz quantum gravity \cite{agsw}. 
However, in
contrast to the case of two-dimensional DT, 
it has been difficult to obtain solvable models of two-dimensional CDT coupled 
to field theories. The only analytically solvable example of an explicit field theory 
system coupled to gravity is provided by CDT coupled to gauge fields \cite{ai}, but 
two-dimensional gauge field theories are mainly topological, so the 
systems obtained are very simple from a matter perspective.

Computer simulations indicate that for unitary conformal field theories 
with central charge $c \leq 1$ the coupling between matter and 
geometry is weak \cite{aal,aalp} with
the critical exponents of both the  matter 
theories and the geometry apparently unchanged. This is in sharp contrast to 
the DT situation, where both matter and geometric exponents are shifted
relative to the matter exponents in flat spacetime and the geometric 
exponents in 2d Liouville gravity without matter. According 
to the computer simulations the situation changes when the central 
charge $c$ of the matter fields coupled to CDT is larger than one indicating  
that the coupling to geometry then becomes strong \cite{aal1,krakow}.

Thus it was interesting and surprising when it was shown that 
restricted dimer systems coupled to CDT could be solved
analytically and seemingly led to 
a change of the critical exponents of the geometry \cite{atkin-zohren,aggs}.
It is well known that the  hard dimer model on a regular two-dimensional 
lattice exhibits critical behaviour for a certain negative value of 
the fugacity, and that this critical system can be associated with
a (2,5) minimal non-unitary field theory having central 
charge $c=-22/5$. In fact, it can be identified via the high temperature
expansion of the Ising model in an imaginary magnetic field with the Lee-Yang 
edge singularity. In \cite{staudacher}
it was shown that a similar identification of a critical hard dimer model
with the Lee-Yang edge singularity can be made in the DT case
and the critical exponents can be calculated. One finds again a  
non-trivial interaction between geometry and matter, but it is weaker
than the interaction between the unitary models and geometry. This is 
in accordance with the expectation that when $c\to -\infty$ matter and gravity decouple. 
Thus the change in critical
geometric properties  found in \cite{atkin-zohren,aggs} when coupling the
CDT model to  some classes of restricted dimers is puzzling: for unitary 
models with $0<c\leq 1$ we have a weak coupling and no change 
in critical geometric properties of the geometry, as mentioned above. 
Naively we would expect the dimer systems at criticality to correspond
to non-unitary field theories with central charge $c <0$ and thus an
even weaker coupling, by analogy to the DT systems. This has motivated us 
to take a closer look at the model  proposed in \cite{atkin-zohren}
(hereafter called the AZ model). As we will report below the model is more subtle 
than anticipated in \cite{atkin-zohren}.

The rest of this article is organized as follows. In Sec.\ 2 we define
a generalized AZ model and discuss its basic properties. 
In Sec.\ 3 we consider the two-point function in the AZ model in a grand canonical 
setting and we use it to calculate the global Hausdorff dimension
$d_H$. Sec.\ 4 addresses the calculation of the so-called 
local Hausdorff dimension $d_h$ in a microcanonical setting. We find somewhat surprisingly 
that $d_h=d_H=3$.
In Sec.\ 5 we show that this result is very special and probably not 
representative for an unrestricted hard dimer model coupled to CDT.
We do this by analyzing in some detail the extended AZ model which allows more general dimer configurations while remaining solvable. The AZ model corresponds to one particular point in the phase boundary of this generalized model and we show that it is the only point at which the Hausdorff dimensions assume the value $3$, while the values at other points are either $d_H=3/2$ and $d_h=1$ or $d_H=d_h=2$. Sec.\ 6 contains a discussion of the results and arguments in favour of viewing $d_H=3/2$ as correct also 
for the  unrestricted dimer model coupled to CDT, i.e.\ for a $c=-22/5$
conformal field theory coupled to 2d Ho\v{r}ava-Lifshitz gravity. Finally, the Appendix discusses the general conditions responsible for the special features of the AZ model.

\section{A restricted dimer model, basic properties}\label{sec:2}

\begin{figure}[t]
\centerline{\scalebox{0.4}{\includegraphics{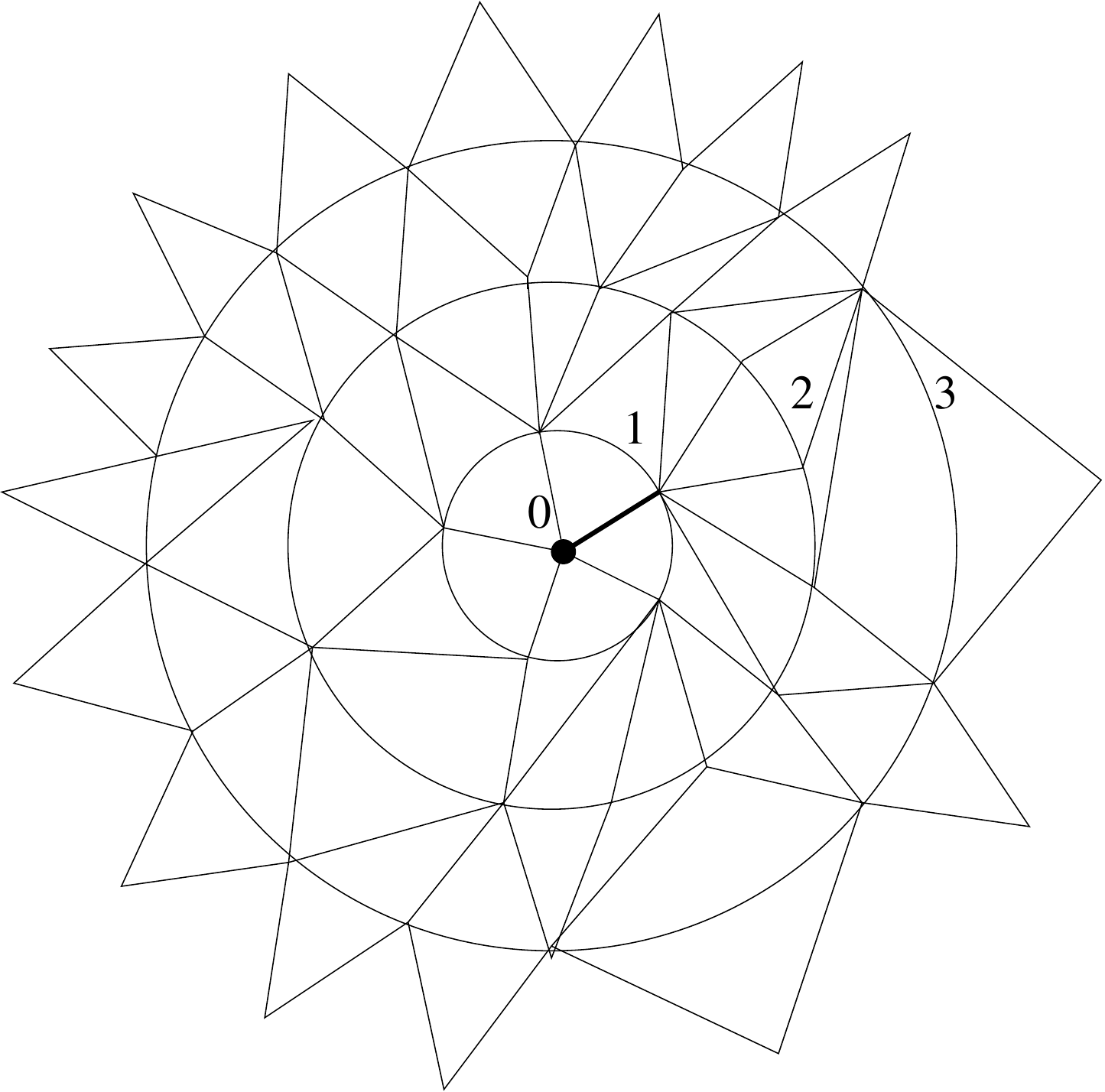}}}
%\centerline{\scalebox{0.4}{\includegraphics{BPandCDT.pdf}}}
\caption{A causal triangulation.}
\label{figCT}
\end{figure}

We consider an extension of the dimer system on random causal triangulations first introduced in \cite{atkin-zohren}. 
A finite causal triangulation $T$ of the planar disc $\cal D$, is constructed as shown in Fig \ref{figCT}. $T$  is 
the union of a central disc $\Sigma_0$ having \emph{central vertex} $v_0$ and 
boundary circle $S_1$, and a sequence of annuli (or time slices) $\Sigma_k,\,k>1,$ such that $\Sigma_k$ is  
bounded by circles $S_{k-1}$ and $S_k$. For $k\geq 1$, $\Sigma_k$ is triangulated by a circular array of triangles 
each of which contains either one vertex in $S_{k-1}$ and two vertices in $S_k$, called a 
\emph{backward directed triangle}, or  two vertices in 
$S_{k-1}$ and one vertex in $S_k$, called a \emph{forward directed triangle}.  $\Sigma_0$ 
is triangulated by a sequence of triangles sharing the central vertex,  which we define to be backward directed. 
Edges contained in one $S_k$ are called \emph{horizontal edges}.    By convention we adjoin a 
forward directed triangle to each of the outermost horizontal edges (see Fig \ref{figCT}) so that all horizontal 
edges are shared by a forward and a backward directed triangle. 
We assume in the following that the edges/vertices in $S_k$ and triangles in $\Sigma_k$ are ordered clockwise, and it 
is convenient to assume also that one of the edges emanating from the central vertex is marked. 

\begin{figure}
\centerline{\scalebox{0.6}{\includegraphics{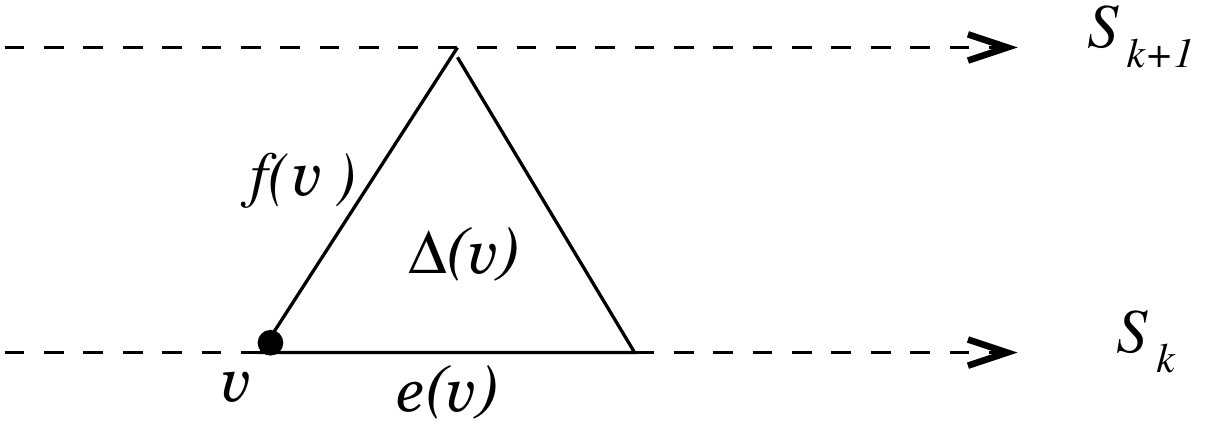} } }
\caption{Labelling the edges of a forward directed triangle. The arrows indicate the clockwise direction on the 
time slices.}
\label{bijection}
\end{figure}

Given a vertex $v\neq v_0$ in $S_k$  we denote by $e(v)$ the  horizontal edge  in $T$ 
emanating in positive clockwise direction from $v$, by $\Delta(v)$ the forward directed triangle containing 
$e(v)$ in its boundary, and by $f(v)$ the non-horizontal edge in $\Delta(v)$ emanating from $v$, see Fig \ref{bijection}. 
Moreover, the \emph{forward degree} $\sigma_f(v)$ and the \emph{backward degree} 
$\sigma_b(v)$ of $v$ are defined as the number of neighbours of $v$ in $S_{k+1}$ and  $S_{k-1}$, respectively. Note 
that $\sigma_f(v), \sigma_b(v) \geq 1$ and that $\sigma_f(v)=1$ if and only if $e(v)$ separates two forward directed 
triangles. 
 
Given a causal triangulation $T$, let $\tilde T$ denote its dual graph. 
A \emph{restricted dimer configuration} $D$ on $\tilde T$ is a set of edges in $\tilde T$ 
fulfilling 
\begin{enumerate}
\item[a)] no pair of edges in $D$ share a vertex in $\tilde T$;
\item[b)] edges in $\tilde T$ dual to edges in $T$ that separate two backward 
directed triangles are not admissible in~$D$;
\item[c)] if $v\in S_k,\, k>0,$ is a vertex with $\sigma_f(v)=1$, then $f(v)$ is not dual to a dimer if either $\sigma_b(v)>1$ or if $\sigma_b(v)=1$ and the successor $w$ to $v$ in $S_k$ has $\sigma_b(w)>1$. 
 \end{enumerate}
Here, condition a) is the standard requirement specifying a dimer configuration, while b) and c) represent further technical 
conditions allowing an exact solution by utilizing a mapping onto labelled trees as demonstrated below.

\begin{figure}[t]
\centerline{\scalebox{0.6}{\includegraphics{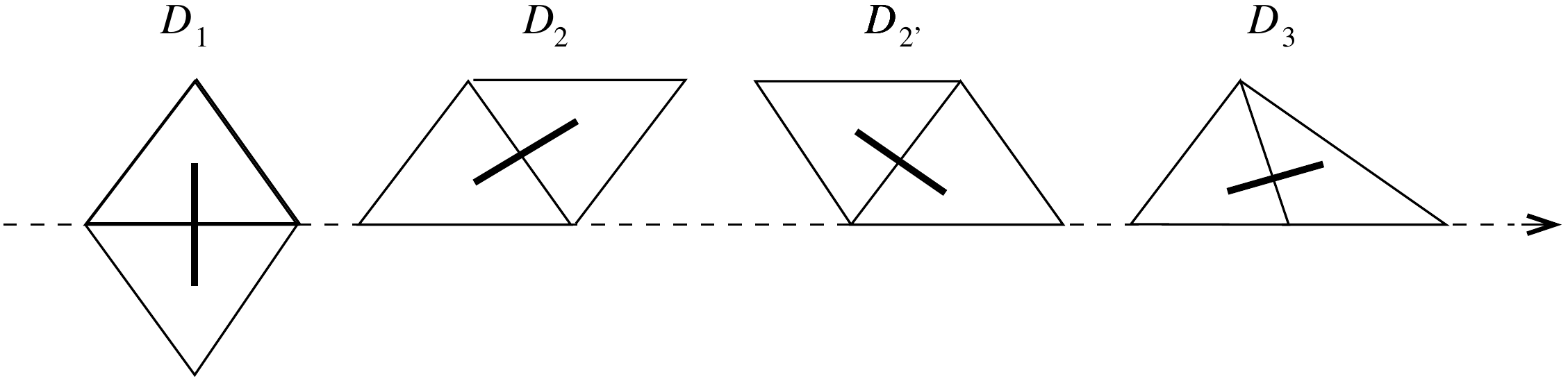}}}
\caption{Dimer Types. The arrow indicates the clockwise direction on a time slice.}
\label{figDimerTypes}
\end{figure}

The possible dimer types are illustrated in Fig \ref{figDimerTypes}. Dimers dual to horizontal edges we call type 1, 
while those   shared by a forward and a backward directed triangle in the same time slice we call types 2 and 2' 
respectively depending on whether or not the backward triangle precedes the forward triangle w.r.t. clockwise ordering. 
Type 3 dimers are those dual to edges shared by two forward directed triangles. 
Finally we denote by $D_i$ the set of edges of type $i$ in $D$ so that 
 $D=D_1\cup D_2 \cup D_{2'}\cup D_3$.

The grand canonical ensemble we are interested in consists of elements $(T,D)$ specified by a causal triangulation 
$T$ and an admissible dimer configuration $D$ on $\tilde T$. With the three types of  dimers we associate 
fugacities $\xi_1,\xi_2,\xi_{2'},\xi_3$ and define the partition function
\beq
Z(\xi_1,\xi_2,\xi_{2'},\xi_3;g) = \sum_{(T,D)} g^{|T|/2}\xi_1^{|D_1|}\xi_2^{|D_2|}\xi_{2'}^{|D_{2'}|}\xi_3^{|D_3|}\,,
\eeq
where $|A|$ denotes the number of elements in a set $A$. It is easy to show that $Z(\xi_1,\xi_2,\xi_{2'},\xi_3;g)$ is 
well defined  for any fixed values of $\xi_1,\xi_2,\xi_{2'},\xi_3$ provided that $|g|$ is sufficiently small.  
 
It is straightforward to see \cite{atkin-zohren} that $Z$ is a function of $\xi_2+\xi_{2'}$ for fixed $g$ and $\xi_1,\xi_3$. Therefore, we shall 
henceforth set $\xi_{2'}=0$, i.e. we further restrict dimer configurations such that $D_{2'}=\emptyset$, and drop $\xi_2'$ 
from the notation and write $\vxi = (\xi_1,\xi_2,\xi_3)$. Note that for $\xi_2=\xi_3=0$ we have 
$Z(\xi_1,0,0;g)= Z(0,0,0;g(1+\xi_1))$, since the number of triangles in a causal triangulation equals twice the number 
of horizontal edges and because horizontal dimers are mutually independent in the absence of non-horizontal dimers.   

In order to determine the analyticity properties of $Z$ we shall, as mentioned, exploit a correspondence between 
admissible pairs $(T,D)$ and certain labelled trees which we now explain by slightly generalising a 
construction given in \cite{atkin-zohren}. 
From a causal triangulation $T$ one  obtains a planar rooted 
tree $\tau=\beta(T)$ in the following way:
\begin{enumerate}
\item[i)] delete all boundary 
edges (those belonging to the outmost forward directed triangles);
\item[ii)] delete all horizontal edges $e(v)$ and all non-horizontal edges of the 
form $f(v)$ for $v\in T$;
\item[iii)] attach a new root edge to $v_0$ such that the marked edge in $T$ is the rightmost 
edge emanating from $v_0$ in $\beta(T)$. 
\end{enumerate}
It was shown in \cite{DJW1} that $\beta$ yields a bijective correspondence between  
causal triangulations with $2n$ triangles and rooted planar trees with n+2 vertices and  root of order $1$.   Note 
that the vertices of $\beta(T)$ different from the root  are also vertices of $T$ and that from now on when 
referring to a tree  we will denote the vertex next to the root by $v_0$ and call it the \emph{first vertex} of the tree. A  dimer 
configuration $D$ on $\tilde T$ induces the following labelling $\ell$ of the vertices of the tree $\tau=\beta(T)$:
\begin{enumerate}
\item[1)] if $e(v)$  is dual to a dimer in $D$, set $\ell(v)=1$;
\item[2)]  if $f(v)$ is dual to a dimer in $D$ and $\sigma_f(v)>1$, set $\ell(v)=2$;
\item[3)] if  $f(v)$ is dual to a dimer in $D$ and $\sigma_f(v)=1$, set $\ell(v)=3$;
\item[4)] the root is unlabelled;
\item[5)] otherwise, set $\ell(v)=0$.    
\end{enumerate}
The restrictions imposed on $D$ are equivalent to the following constraints on $\ell$:
\begin{itemize}
\item[a)] if a leaf of $\beta(T)$ has label $3$ then its preceding neighbour at same height has label $0$ and 
its successor at same height has label $0$ or $1$;
\item[b)] if a vertex in $\beta(T)$ that is not a leaf has label $2$, then its rightmost decendant does not have 
label $1$;
\item[c)] a leaf of $\beta(T)$ that is the leftmost or the rightmost decendant of its predecessor does not have label $3$.
\end{itemize}

Noting that all edges in $T$ dual to dimers in $D$ are deleted when constructing  $\beta(T)$, it is  
straightforward  to show that the correspondence between pairs $(T,D)$ and 
pairs $(\tau,\ell)$ with $\ell(v_0(\tau))=0$  is bijective.
Note also that the number $\ell_i$ of vertices in $\beta(T)$ with label $i$ equals $|D_i|$ for $i=1,2,3$.   

Consider now a labelled tree $(\tau,\ell)$ as above and assume the vertex $v_0$ next to the root has order $s+1$, 
i.e. $\tau$ has $s$ branches $\tau_1,\ldots,\tau_s$ rooted at $v_0$. Since $\ell(v_0(\tau))=0$ the labellings of the $\tau_i$ induced by the labelling of $\tau$ are independent and the labelling of the first vertex is unrestricted.
%
% The partition function for trees whose first vertex has label $i$ is defined to be
%The labelling $\ell$ induces a labelling
%
%and since $\ell(v_0(\tau))=0$
% is not 
%labelled 
%the labellings of the branches are independent. %and only subject to restrictions a) and b). 
%$For $i=0,1,2$ we set
%Defining
%\beq\label{W}
%W_i(\xi_1,\xi_2,g) = \sum_{(\tau,\ell):\ell(v_0)=i} g^{|\tau|}\xi_1^{\ell_1}\xi_2^{\ell_2} ,\qquad  i=0,1,2.
%\eeq
%where the sum is over all rooted, labelled planar trees with root $v_0$ of order $1$ and unlabelled, and with vertex $v_1$ next to 
%the root labelled by $i$. Here 
%where $|\tau|$ denotes the number of edges in $\tau$.  %. From the preceding remarks 
However, the labellings of the different branches rooted at the vertex $v_0(\tau_i)$ may, depending on its label, 
not be  independent due to the constraints a) and b). 
Now define the  partition function for trees whose first vertex has label $i$ to be
\beq\label{W}
W_i(\vxi;g) = \sum_{(\tau,\ell):\ell(v_0)=i} g^{|\tau|}\xi_1^{\ell_1}\xi_2^{\ell_2}\xi_3^{\ell_3} ,\qquad  i=0,1,2,
\eeq
where $|\tau|$ denotes the number of edges in $\tau$. Then decomposing trees into the root edge and their branches rooted at 
$v_0$ shows that the $W_i$ satisfy the equations
\beq \label{j3} W_i=F_i(W_0,W_1,W_2; \vxi; g)\,,\eeq
% the 
 %relations
%\bea \label{j3}
%W_0(\xi_1,\xi_2,g) &=& \frac{g}{1-W(\xi_1,\xi_2,g)}\label{j1}\nn
%W_1(\xi_1,\xi_2,g) &=& \xi_1\, W_0(\xi_1,\xi_2,g) \label{j2}\nn
%W_2(\xi_1,\xi_2,g)&=& \xi_2 \,g\,\frac{W_0(\xi_1,\xi_2,g)+W_2(\xi_1,\xi_2,g)}{1-W(\xi_1,\xi_2,g)}\label{j3}
%\eea
where
\begin{eqnarray*}\label{Fdefinitions}
F_0&=&g\,(1-g\xi_3 W_0)H\\
F_1&=& g\,\xi_1(1-g\xi_3 W_0)H\\
F_2&=&g\,\xi_2\left(W_0+(1-g\xi_3 W_0)W_2\right)H 
\end{eqnarray*}
with
\beq\label{H}
 H = \left\{1-(1+g\xi_3) W_0 - W_1- (1-g\xi_3 W_0)W_2\right\}^{-1}\,.
\eeq
It follows from the discussion above that
\beq\label{ZW}
Z(\vxi;g) = g^{-1}W_0(\vxi;g) - 1\,,
\eeq
and that non-analytic behaviour of $Z(\vxi;g)$ as a function of $g$ at a critical point $g_c(\vxi)$, of the form  
$$
Z_c - Z(\vxi;g) \sim  (g_c-g)^\alpha\,,
$$
for some $0<\alpha<1$, occurs if and only if $W_0$ exhibits the same behaviour 
\beq\label{gc}
W_{0c} - W_0(\vxi;g) \sim (g_c-g)^\alpha\,.
\eeq
Here and in the following the notation $a\sim b$ is used to indicate that there exist constants $c_1, c_2>0$ such 
that $c_1\, a\leq b\leq c_2\, a$.
By eliminating $W_1, W_2$ from  \eqref{j3} we obtain
\beq\label{W0}
(\xi_1+g\xi_3)\xi_2 W_0^3 -(1+\xi_1+\xi_2 +g\xi_3 + g^2\xi_2\xi_3) W_0^2 + (1+g\xi_2 + g^2\xi_3)W_0 -g =0
\eeq
which, for  fixed $\vxi$,  determines $W_0$ as the unique 
root vanishing and analytic at $g=0$. For $\xi_1,\xi_2,\xi_3\geq 0$ the Taylor expansion of $W_0$ obtained from \eqref{W} has positive coefficients and hence its radius of convergence $g_c>0$ is a singularity of $W_0$. This is  a property of $W_0$
that persists, as we shall see, in a larger range $S$ of couplings $\xi_1,\xi_2,\xi_3$ that are not necessarily positive. For generic $\vxi$ in this range, $W_{0c}$ is 
a double root of \eqref{W0} at $g=g_c$
and  $g_c$ is a square root branch point of $W_0$ as a function of $g$. If  $\vxi\in S$  is such that $W_{0c}$ is a 
triple root of \eqref{W0} at $g=g_c$ , i.e. $\alpha =1/3$ in \eqref{gc}, then
$(\vxi;g_c)$ is a multicritical point which we denote by $(\vxi_c;g_c)$.
 % \begin{figure}
%\centerline{\scalebox{0.6}{\includegraphics{QuadraticWithAxes.pdf}\includegraphics{CubicWithAxes.pdf}}}
%\caption{Graphical representation of $g(\xi,\xi,0,W_0)$ a) for 
%$\xi >\xi_c$ and b) for $\xi=\xi_c$.}
%\label{figj2}
%\end{figure}

 The condition that $g_c$ be a double root is obtained by differentiating \eqref{W0}  w.r.t $W_0$, so that     
%Writing 
%\beq\label{j8}
%g(\xi_1,\xi_2,W_0) = \frac{W_0(\xi_1\xi_2 W_0^2-(1+\xi_1+\xi_2) W_0+1)}{1-\xi_2 W_0}\,,
%\eeq
the critical coupling $g_c$ and the corresponding value $W_{0c}$ of $W_0$ satisfy 
\beq\label{j9} 
3(\xi_1+g_c\xi_3)\xi_2 (W_{0c})^2-2(1+\xi_1+\xi_2+g_c\xi_3+ g^2\xi_2\xi_3)W_{0c}+(1+g_c\xi_2 + g_c^2\xi_3)=0.
\eeq
Using this in \eqref{W0} yields 
\beq\label{j10}
(1+\xi_1+\xi_2+g_c\xi_3+ g_c^2\xi_2\xi_3)(W_{0c})^2-2(1+g_c\xi_2 + g_c^2\xi_3)W_{0c} +3g_c=0.
\eeq
and $g_c$ and $W_{0c}$ are determined as functions of $\vxi$ by  \rf{j9} and \rf{j10}.

Multicritical points additionally satisfy 
\beq\label{j11}
3(\xi_{1c}+g_c\xi_{3c})\xi_{2c} W_{0c}-(1+\xi_{1c}+\xi_{2c}+g_c\xi_{3c}+ g_c^2\xi_{2c}\xi_{3c})=0\,. 
\eeq
%Hence 
%\beq\label{j11a}
%W_{0c} = \frac{1+\xi_{1c}+\xi_{2c}}{3\xi_{1c}\xi_{2c}}= \frac{1+\xi_{2c}g_c}{1+\xi_1+\xi_2}
%\eeq
%t a multicritical point $(\xi_{1c},\xi_{2c},\xi_{3c},g_c)$. 
The existence of such multicritical points can be established by e.g. 
setting $\xi_1=\xi$, $\xi_2=\kp \xi$ and $\xi_3=0$, where $\kp>0$. The value
$\kp=2$ corresponds to the AZ model considered in \cite{atkin-zohren}. However
the results are universal for $\kp>0$ and we will 
denote all these models as AZ models, and explicitly 
perform the calculations for $\kp=1$ (except in footnote \ref{ftn1} where
we discuss the situation of an arbitrary real value of $\kp$).
For $\kp=1$ we  use \eqref{j9}, 
\eqref{j10} and \eqref{j11} to obtain the equation 
\beq\label{jjx}
\xi_c^3 + 24 \xi_c^2 + 3\xi_c -1 =0\,
\eeq
for the critical value of $\xi$.
This polynomial has one positive root and two negative roots and a 
closer analysis shows that the largest negative 
root  $\xi_c \approx -0.278$ corresponds to a 
multicritical point $(\xi_c,\xi_c,0;g_c)$, 
while $g_c$ is a square root singularity 
of $W_0$ for $\xi>\xi_c$ (for a discussion of the situation for a 
general value of $\kp$ see footnote \ref{ftn1}). 

%In sections \ref{sec:3} and \ref{sec:4} below we discuss the case  $\xi_1=\xi_2=\xi$ and $\xi_3=0$ in more detail before  returning to the general case in section \ref{sec:5}.

%The graphs of $g$ as a function of $W_0$ in the generic case and the multicritical cases are shown in 
%Fig.\ \ref{figj2}. 

\section{The two-point function for $\xi_1=\xi_2=\xi, \xi_3=0$ }\label{sec:3}

%Having described the singular behaviour of the partition functions 

We now consider  the fractal behaviour of triangulations 
and the corresponding labelled trees close to the critical point for the AZ model. The Hausdorff dimension in the grand canonical ensemble is determined 
by the decay rate of the two point function  \cite{book}.%  to which we now turn. 
%As already mentioned, we assume in this section that $\xi_1=\xi_2=\xi, \xi_3=0$.  

\begin{figure}[t]
\centerline{\scalebox{0.6}{\includegraphics{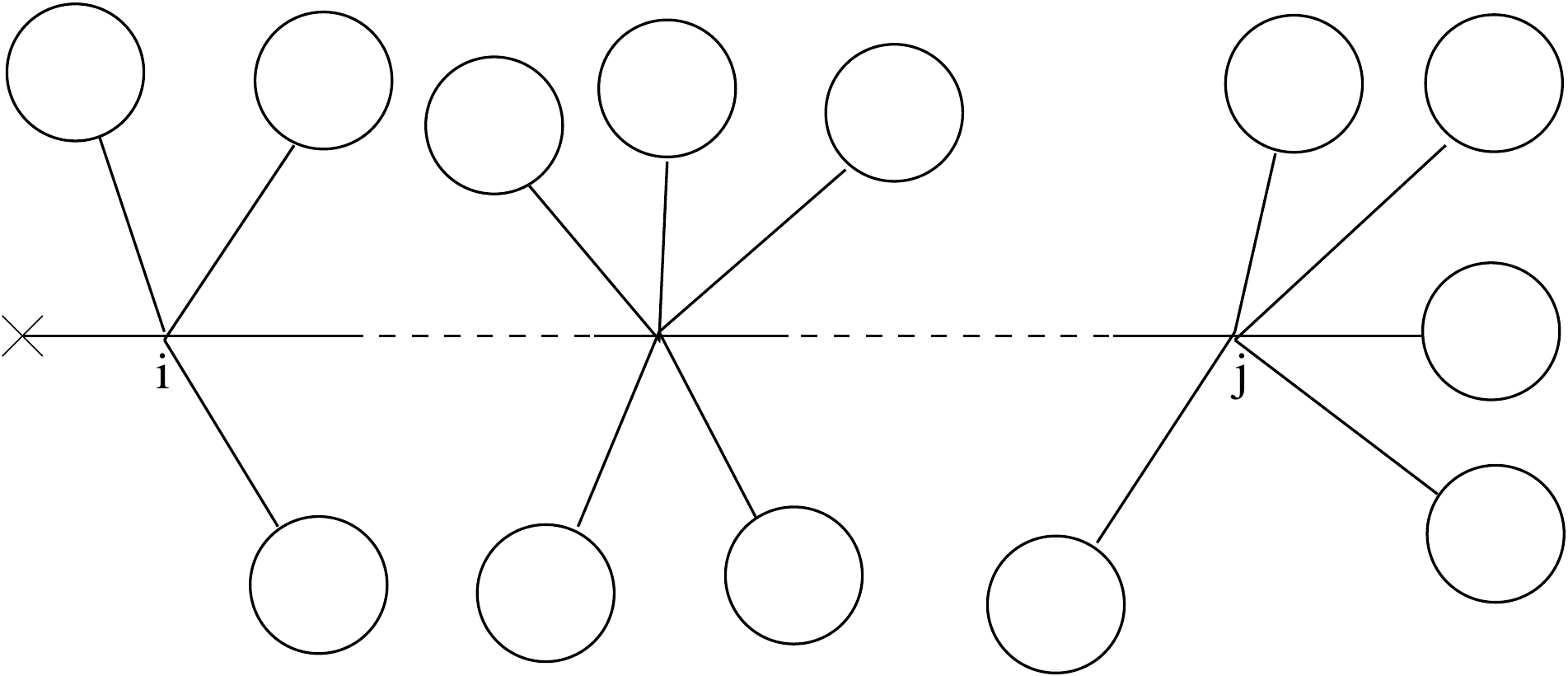}}}
%\centerline{\scalebox{0.5}{\includegraphics{dual.pdf}}}
\caption{Graphical illustration of $\mathbb G(\xi,g;r)_{ij}$.}
\label{figj1}
\end{figure}

Concentrating first on trees define  a marked 
labelled tree  to be a triple $(v,\tau,\ell)$ where $(\tau,\ell)$ is a labelled tree as above and $v$ is a vertex in 
$\tau$ different from the root and the first vertex $v_0$. By $d(v)$ we denote the graph distance from the root to $v$. 
The  two-point function  $\mathbb G(\xi,g;r)$ is  defined by 
$$
(\mathbb G(\xi, g;r))_{ij} = W_j^{-1}\sum_{(v,\tau,\ell):\ell(v_0)=i,\ell(v)=j,d(v)=r+1} g^{|\tau|}\xi^{\ell_1+\ell_2} \,,
$$
for  $i,j\in\{0,1,2\}$ and $r\geq 1$.
%
% For $i,j\in\{0,1,2\}$ the two-point function  $G_{ij}(\xi,g;r)$ is defined by 
%$$
%G_{ij}(\xi, g;r) = W_j^{-1}\sum_{(v,\tau,\ell):\ell(v_0)=i,\ell(v)=j,d(v)=r+1} g^{|\tau|}\xi^{|D_1|+|D_2|} \,,
%$$
%for $r\geq 1$. 
A schematic illustration of the two-point function is shown in Fig.\ \ref{figj1}, from which it follows by standard arguments 
and considerations similar to those leading to \eqref{j3} that 
  
\beq\label{j4}
{\mathbb G}(\xi,g;r) = \mathbb T(\xi,g)^r\,,
\eeq
where
\beq 
(\mathbb T(\xi,g))_{ij} =\frac{\partial F_i}{\partial W_j}
\eeq
which after some simplification gives
\bea\label{j5}
\mathbb T (\xi,g)
 %&=& %g^{-1}W_0^2
%\begin{pmatrix}
%1 & 1 & 1 \\
%\xi &\xi & \xi \\
%\xi(1-W_1) & \xi(W_0 + W_2) & \xi (1-W_1) 
%\end{pmatrix} \nonumber\\
&=& g^{-1}W_0^2
\begin{pmatrix}
1 & 1 & 1 \\
\xi &\xi & \xi \\
\xi(1-\xi W_0) & {\xi W_0}(1- \xi W_0)^{-1} & \xi (1 -\xi W_0) \end{pmatrix}\;.
\eea

%In order to determine the decay of $G_{ij}$ as a function of $r$ we proceed to write  $\mathbb G$ in Jordan normal form. 
%$\hG = \hP \hJ \hP^{-1}$, where $\hP$ is a real matrix. Thus we have 
%\beq\label{j6}
%\hG(r) = \hP \hJ^r \hP^{-1}.
%\eeq
%We now determine $\hJ$.

%From  eqs.\ \rf{j1}-\rf{j3} we have:
%\beq\label{j7}
%\xi^2 Z_0^3 -(1+2\xi)Z_0^2+(1+\xi g)Z_0 -g=0.
%\eeq
%This equation determines $Z_0(\xi,g)$. We can also write 
%\beq\label{j8}
%g(\xi,Z_0) = \frac{Z_0(\xi^2Z_0^2-(1+2\xi Z_0)+1)}{1-\xi Z_0}.
%\eeq%
%
%A critical point is determined by :
%\beq\label{j9}
%\frac{\prt g}{\prt Z_0} = 0,~~~~{\rm i.e.}~~~~ 
%3\xi^2 (Z_0^c)^2-2(1+2\xi)Z^c_0+(1+\xi g_c)=0.
%\eeq
%Using this in \rf{j7} we have (at the critical point):
%\beq\label{j10}
%(1+2\xi)(Z_0^c)^2-2(1+\xi g_c)Z_0^2 +3g_c=0.
%\eeq
%Eqs. \rf{j9}-\rf{j10} determine $g_c(\xi)$ and $Z_0^c(\xi)$.%
%
%The multicritical point is determined by the additional requirement:
%\beq\label{j11}
% \frac{\prt^2 g}{\prt Z_0^2} = 0,~~~~{\rm i.e.}~~~~~ 
%3\xi_m^2 Z^m_0-(1+2\xi_m)=0, 
%\eeq
%i.e.\
%\beq\label{j11a}
%Z_0^m = \frac{1+2\xi_m}{3\xi_m^2}
%= \frac{1+\xi_m g_m}{3\xi_m^2}= \frac{3g_m}{1+\xi_mg_m}.
%\eeq
%where we use sub- and superscript ``m'' for multicritical.
%
%The graph for $g(\xi,Z_0)$ asa function of $Z_0$ 
%for $0 \leq \xi \leq \xi_m$ and for $\xi= \xi_m$ is shown
%in Fig.\ \ref{figj2}.
%\begin{figure}[t]
%\centerline{\scalebox{0.4}{\includegraphics{gZcurve.eps}}}
%\caption{The graphical representation $g(\xi,Z_0)$ for 
%$\xi >\xi_m$ and for $\xi_m$.}
%\label{figj2}
%\end{figure}

Clearly, the matrix $\mathbb T$ has one eigenvalue $\lam_3=0$. The two
other eigenvalues, $\lam_1$ and $\lam_2$, are solutions of the characteristic equation
\beq\label{j12}
\lam^2 -\Big(-1+(1+\xi g)\frac{W_0}{g}\Big)\, \lam +\frac{\xi^2W_0^3}{g}=0.
\eeq
which we rewrite as
\beq\label{j13}
\Big(\lam-1\Big)\Big(\lam -\frac{\xi^2W_0^3}{g}\Big) +
\lambda(1-\xi W_0) \,\frac{W_0}{g}\,\frac{\prt g}{\prt W_0}=0.
\eeq 
On the critical line 
$(W_{0c}(\xi),g_c(\xi)),\,\xi\geq \xi_c,$  
the last term vanishes and 
\beq\label{j14}
\lam_1=1,~~~\lam_2 = \frac{\xi^2W_{0c}(\xi)^3}{g_c(\xi)}  ~~~~{\rm for}~~\xi \geq \xi_c.
\eeq
In particular, for $\xi=0$ we have $W_{0c}=1/2$, $g_c=1/4$ and $\lam_2=0$. As $\xi$
decreases from zero to $\xi_c \approx -0.278$ we find that 
%\footnote{$\xi_m \approx -0.278$,  $Z_0^m\approx 1.91$, $g_m \approx 0.54$}, 
$\lam_2$  increases monotonically to
$\lam_{2c} = 1$, where the value $1$ is a simple consequence of  \eqref{W0} and \rf{j11}. 

For fixed $\xi>\xi_c$ the two eigenvalues are real for $\Delta g= g-g_c$ small enough and the  
dominant eigenvalue is  $\lam_1$  approaching $1$ as $g\to g_c$. Hence, we get in this case
$$
\mathbb G(\xi,g;r) =  e^{-m(g)r + o(r)}
$$
as $r\to\infty$, where
$$
m(g) \sim \Delta \lam_1 = 1-\lam_1\,.
$$
At the critical line for $\xi > \xi_c$ we have $g'(W_{0})=0$ and 
$g''(W_0)\neq 0$, i.e.\ $\alpha = 1/2$ in eq.\ \eqref{gc}. Thus  
we obtain from \rf{j13} 
\beq\label{j16}
\Del\lam_1 \sim \frac{\prt g}{\prt W_0}\sim  |\Del g|^\oh.
\eeq
Hence, the two-point functions decay exponentially with rate
$$
m(g) \sim  |\Del g|^{\frac 12}\,. 
$$
This shows that for the labelled trees with $\xi>\xi_c$ the \emph{global  Hausdorff dimension}, defined as the inverse of 
the critical exponent of $m(g)$ (see e.g. \cite{book}), is $d_H=2$. 

At the multicritical point $(\xi_c,g_c)$, on the other hand, we have $g'(W_0) = g''(W_0)=0$ and $\alpha = 1/3$ in \eqref{gc}. 
% such that
%\beq\label{j15a}
% \Del g \sim (\Del W_0)^3,~~~\frac{\prt g}{\prt W_0}\sim (\Del W_0)^2.
%\eeq
%where $\Del W_0 = W_0-W_o^c$.
%In the simpler multicritical BP model discussed by Ambjorn et al.
%(\cite{amb}) one has an equation similar to \rf{j13}, but with
%the simplicifation that $\lam_2^m<1$. Thus even at the multicritical 
%point there is only one dominant eigenvalue, and one finds
%using \rf{j15a}: 
%\beq\label{j17}
% \Del\lam_1 \sim \frac{\prt g}{\prt Z_0}\sim (\Del Z_0)^2 \sim |\Del g|^{2/3}.
%\eeq
%This implies that we in this model has the exponential decay 
%$e^{-c \, |\Del g|^{2/3} r}$ and thus the ``global'' 
%Hausdorff dimension $d_H=3/2$. 
%
%Since $\lam_2=\xi^2Z_0^3/g\to 1$ as $g \to g_c$ there are two dominant eigenvalues $\lam_1$ and $\lam_2$. 
Using this and \eqref{j11} gives
\beq\label{j18}
(1-\xi_c W_0) \frac{\prt g}{\prt W_0} = 3 \xi_c^2 (\Del W_0)^2 + \xi_c\Del g\,,
\eeq
where $\Del W_0 = W_0-W_{0c}$. Inserting this expression into \eqref{j13} and setting $\Del\lam = 1-\lam$ now gives\footnote{\label{ftn1} One can 
repeat the calculations leading to \rf{j19} for the general assignment
$\xi_1=\xi$, $\xi_2= \kp \xi$, $\xi_3=0$, $\kp >0$, and 
 find that \rf{j19} is independent of $\kp$. However 
the values of $W_{0c}$, $g_c$ and $\xi_c$ depend on $\kp$. What is 
important is the existence of a multicritical point $\xi_c$.
For $\kp=1$ this was ensured by eq.\ \rf{jjx}. The equation for a 
general $\kp$ is
$$
-b^3 \xi_c^3 +3(9-b^2)\xi_c^2 -3b\xi_c -1=0,~~~\kp=b+2.
$$
For $\kp >0$ the largest real negative root corresponds to 
the multicritical point, precisely as in \rf{jjx}.  
Interestingly, for the original AZ value
$b=0$ the equation simplifies to a trivial second order equation (it is the 
point where the equation changes from having two negative and one 
positive solution to one negative and two positive solutions, the third root
moving to $-\infty$ for $b\to 0^-$ and to $\infty$ for $b\to 0^+$). However,
this has no consequences for the discussion of multicriticality. 
For $\kp <0$ there is no negative real solution.}  
\beq\label{j19}
(\Del \lam)^2 + \frac{3}{W_{0c}} \, \Del \lam \,\Del W_0 +  
\frac{3}{(W_{0c})^2}\, (\Del W_0)^2 = O(\Del W_0^3),
\eeq
and hence 
\beq\label{j20}
\Del \lam = \frac{3}{2 W_{0c}}  |\Del W_0|(1\pm i /\sqrt{3}) + O(\Del W_0^2).
\eeq
In particular, $\lam_1 = \overline\lam_2$ is complex for $g< g_c$ in this case and we conclude that %$G_{ij}(r)$
   $\mathbb G(\xi,g;r)$ decays  
%
%We can finally write down the Jordan normal form of $\mathbb G$ as
%\beq\label{j21}
%\mathbb J = 
%\begin{pmatrix}
%1-|\Del Z_0|c    & c\sqrt{3}|\Del Z_0|& 0\\
%-c \sqrt{3}|\Del Z_0| & 1-c|\Del Z_0| & 0\\
%0&0&0
%\end{pmatrix}
%\eeq
%from which we obtain an exponential fall off $e^{-c r |\Del g|^{1/3}}$
exponentially (dressed with oscillating factors) with decay rate 
$$
m(g) \sim \mbox{Re}\,\Del\lam   \sim  |\Del W_0|   \sim |\Del g|^{\frac 13}\,.
$$
This yields the value $d_H=3$  for the global Hausdorff dimension at $\xi=\xi_c$.

Returning to the  dimer model on causal dynamical triangulations the two-point function $G(\xi,g;r)$ is defined in 
the same manner as for trees by marking a vertex $v$ at distance $d(r)$ from the central vertex of the triangulation $T$ and setting
$$
G(\xi,g;r) = \sum_{(v,T,D):d(v)=r} g^{|T|/2}\xi^{|D_1|+|D_2|}\,.
$$
Using the mapping $\beta$ between the dimer model and the labelled tree model we obtain
$$
G(\xi,g;r) = g^{-1}\sum_j \mathbb G_{0j}(\xi,g;r) W_j(g)\,.
$$
In particular, $G(\xi,g;r)$ has the same exponential decay rate as the two-point functions $\mathbb G(\xi,g;r)_{ij}$, and hence 
the global Hausdorff dimension of the dimer model coincides with that of the labelled tree model, i.e. $d_H=2$ for $\xi>\xi_c$ and 
$d_H=3$ for $\xi=\xi_c$.

%It is interesting to compare the result of this section with the simpler multicritical tree model discussed in
%\cite{amb} and reviewed briefly in Appendix 1. In this case an equation similar to \rf{j13} is obtained, but  $\lam_{2c}<1$. Thus even% at the multicritical 
%point there is only one dominant eigenvalue, and one finds,
%using \rf{j16}, that   
%\beq\label{j17}
% \Del\lam_1 \sim \frac{\prt g}{\prt W_0}\sim (\Del W_0)^2 \sim |\Del g|^{2/3}.
%\eeq
%which implies that the Hausdorff dimension of the model is $d_H=3/2$ for $\xi=\xi^c$. 

\bigskip

\section{The infinite size limit}\label{sec:4}

In this section we consider an alternative notion of Hausdorff dimension for the labelled tree model by considering only critical 
trees, i.e. we shall evaluate the infinite size limit first and express the Hausdorff dimension in terms of volume growth on 
 infinite trees. In order to make this precise, let us  introduce the finite size partition functions $W_{iN}(\vxi)$ by restricting the 
sum in \eqref{W} to trees $\tau$ of fixed size $N$, so that 
$$
W_i(\vxi;g) = \sum_N g^N W_{iN}(\vxi)\,.
$$
The distributions $\mu_{iN}$ of labelled trees of fixed size $N$ are obtained by normalizing the weights defining  $W_{iN}$, i.e.
$$
\mu_{iN}(\tau,\ell) = \frac{1}{W_{iN}(\vxi)} \xi_1^{\ell_1}\xi_2^{\ell_2}\xi_3^{\ell_3}\,.
$$
 Obviously, $\mu_{0N}$ is non-negative whenever $\xi_1,\xi_2,\xi_3\geq 0$ and hence defines a probability distribution. For $\xi_1=\xi_2=\xi$ and $\xi_3=0$ it is not difficult to see that this even holds true for $\xi\geq -\frac 14$, but not for $\xi_c\leq \xi< -\frac 14$. Similar remarks apply to $\mu_{iN}$ up 
to a sign factor. Our aim is to consider limits of the expectations $\la\cdot\ra_{iN}$ with respect to the (signed) distributions $\mu_{iN}$ as $N\to\infty$, for arbitrary values of $\vxi\in S$. 

As a consequence of \eqref{gc} and standard transfer theorems (see e.g. \cite{FS:2009})  the following asymptotic behaviour of $W_{iN}(\vxi)$ for large $N$ holds:
\bea
W_{iN}(\vxi) &=&  \Omega_i N^{-3/2} g_c(\vxi)^{-N}(1+\mbox{O}(\tfrac 1N))\quad\mbox{if $\alpha=1/2$}\,,\label{asymp1}\\
W_{iN}(\vxi) &=&  \Omega_i N^{-4/3} g_c(\vxi)^{-N}(1+\mbox{O}(\tfrac 1N))\quad\mbox{if $\alpha=1/3$}\,,\label{asymp2}
\eea
 where the constants $\Omega_i$ depend on $\vxi$. We note that the relations \eqref{j3} imply
\bea
\Omega_1 &=& \xi_1\Omega_0  \label{Omeg1}\\
\Omega_2 &=& \xi_2 W_{0c}\frac{2-\xi_2 W_{0c}-g_c\xi_3W_{0c}}{(1-\xi_2 W_{0c})^2(1-g_c\xi_3W_c)^2}\,\Omega_0\,.  \label{Omeg2}
\eea

 Using \eqref{asymp1} and \eqref{asymp2} it follows by a straight-forward generalization of arguments given in \cite{BD,DJW1} that 
for any local quantity $A(\tau,\ell)$ depending only on the structure of $(\tau,\ell)$ within a finite distance $R$ from the root of $\tau$, such 
as the volume of the ball $B_R(\tau)$ of radius $R$ centered at the root, the limiting expectation values 
$$\la A\ra_i = \lim_{N\to\infty}\la A\ra_{iN}$$
 exist. 
%the main 
%ingredient for applying the methods of \cite{BD,DJW1} to expectation values is the fact that the exponents $-3/2$ and $-4/3$ 
%in \eqref{asymp1} and \eqref{asymp2} are less than $-1$. 

 \begin{figure}
\centerline{{\includegraphics[scale=0.6,angle=-90]{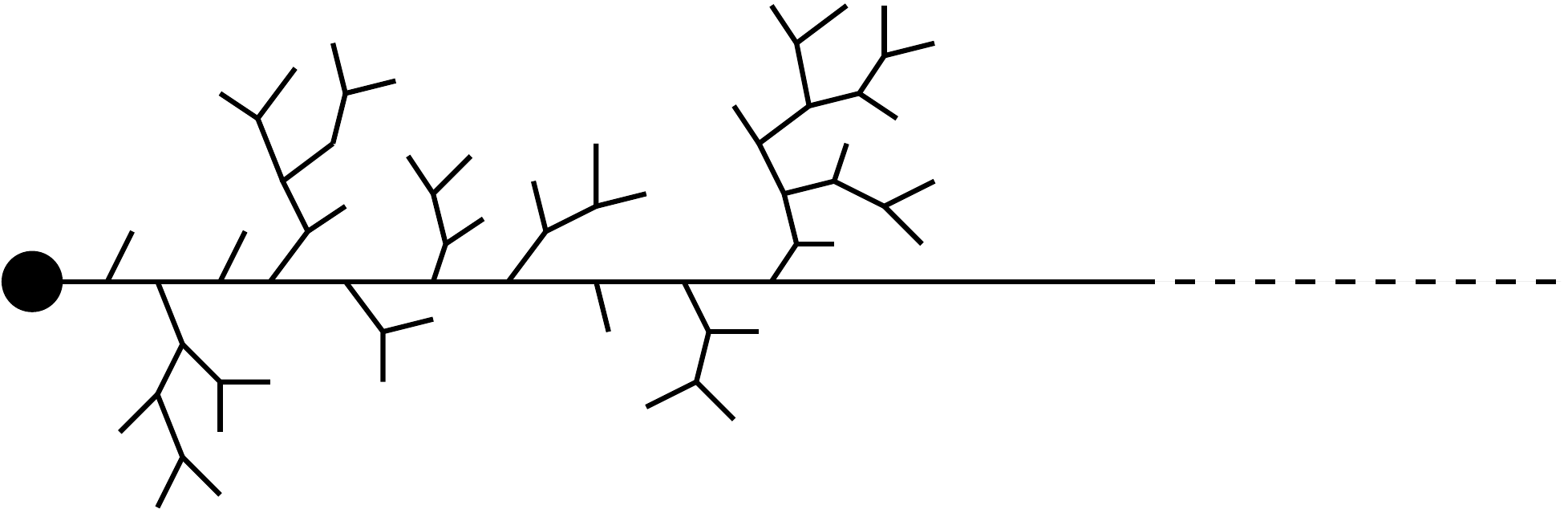}}}
\caption{Finite trees attached to the first few vertices on the spine of an infinite tree.}
\label{SpineTree}
\end{figure}

We next briefly describe how to calculate the limiting expectation values $\langle A\rangle_i$  in terms of infinite labelled trees, further details can be found in \cite{BD,DJW1}. Here an infinite labelled tree means an infinite rooted planar tree with root of order $1$ with a 
labelling respecting the same conditions a)- c) as previously. Moreover, only trees with a single spine, that is an infinite 
self-avoiding path starting at the root, contribute to $\langle A\rangle_i$, see Fig. \ref{SpineTree}.
%--------------------
%-----NB! THIS FIGURE IS MISSING, see Fig. 5.
%-------------------- 
The spine vertices of an infinite labelled 
tree $L$ will be denoted by $r, u_1, u_2, u_3,\dots$, ordered by increasing distance from the root $r$. Thus $L$ is obtained 
by grafting finite labelled trees with root of order $1$, called branches, at the spine 
vertices $u_i$ on both sides of the spine.  Considering only the finite part $r,u_1, u_2,\dots, u_N$ of the spine one of the branches 
rooted at $u_N$ is infinite and all other branches are finite. We denote by $L(N)$ the finite labelled tree obtained by removing 
the infinite branch at $u_N$ except the edge $u_Nu_{N+1}$ and consider $L(N)$ as a finite labelled tree with a finite 
spine $r, u_1, u_2,\dots, u_{N+1}$ both of whose end vertices have order $1$. By  ${\cal A}(L_0)$ we denote the set of all 
infinite labelled trees such that $L(N)$ equals a fixed finite labelled tree $L_0$ with a distinguished spine $r, u_1, u_2,\dots, u_{N+1}$.
With this notation the limiting weight of the set ${\cal A}(L_0)$ equals
\beq\label{mui}
\mu_i({\cal A}(L_0)) = \Omega_i^{-1}\Omega_{\ell_{N+1}} \rho_i(L_0)\,,
\eeq
where $\rho_i(L_0)$ is the grand canonical weight of $L_0=(\tau_0,\ell_0)$ at the critical point $(\vxi;g_c)$ given by 
\beq\label{rhoi}
\rho_i(\tau_0,\ell_0) = \delta_{\ell_0(u_1),i} g_c^{|\tau_0|-1} \xi_1^{\ell_{01}-\delta_{\ell_0(u_{N+1}),1}}\xi_2^{\ell_{02}-\delta_{\ell_0(u_{N+1}),2}}\xi_3^{\ell_{03}} \,.
\eeq

The information contained in  \eqref{mui} and \eqref{rhoi} suffices to calculate $\langle A\rangle_i$ for any local quantity $A$. 
We now proceed to calculate $\langle |B_R|\rangle_i$, where $|B_R(L)|$ denotes the size of $B_R(L)$, i.e. the number of edges 
in $\tau$ whose vertices are at graph distance at most $R$ from the root. The \emph{local Hausdorff dimension} $d_h$ of the 
random tree defined by $\mu_i$ is defined by
\beq\label{haus2}
\langle\; |B_R|\;\rangle_i \; \sim \; R^{d_h}
\eeq
as $R\to\infty$. The purpose of the next two subsections is to evaluate $d_h$ in the case $\xi_3=0$ and demonstrate that its value 
coincides with $d_H$ as found in Section 3.%above in this case.

\subsection{Volume of a finite tree for $\xi_1=\xi_2=\xi$, $\xi_3=0$}

We let $H_i^R$ denote the (unnormalized) expectation value of the number 
of vertices $h_R(\tau)$ at distance $R$ from the root 
of a finite tree $\tau$ with label $i$ on its vertex $v_0$ at the critical value $g_c(\xi)$ of the coupling $g$, that is 
\bea
H_i^R = \sum_{(\tau;\ell):\ell(v_0)=i} h_R(\tau) g_c^{|\tau|} \xi^{\ell_1+\ell_2}\,.
\eea
Applying arguments similar to those leading to  \eqref{j3} one obtains, for $i=0,1,2$,
%\bea
%H_0^R &=& g \frac{H_0^{R-1} + H_1^{R-1} + H_2^{R-1}}{(1-(W_{0c} + W_{1c} + W_{2c}))^2}\label{H1}\\
%H_1^R &=& \xi H_0^R \label{H2}\\
%H_2^R &=& g\xi \frac{H_0^{R-1} + H_2^{R-1}}{1-(W_{0c} + W_{1c} + W_{2c})} + g\xi(W_0 + W_2) \frac{H_0^{R-1} + H_1^{R-1} + 
%H_2^{R-1}}{(1-(W_{0c} + W_{1c} + W_{2c}))^2}\label{H3}
%\eea  
\bea
H_i^1 &=& W_{ic}\,,\nn
H_i^R&=&\sum_{j=0}^2\mathbb T(\xi,g_c(\xi))_{ij} H_j^{R-1}\,,\quad R\geq 2,\label{H3}
\eea 
where $\mathbb T$ is given by \eqref{j5}.  $\mathbb T_c = {\mathbb T}(\xi,g_c(\xi))$ has eigenvalues
\beq \lambda_0=0, \quad
\lam_1=1,\quad \lam_2 = \frac{\xi^2W_{0c}(\xi)^3}{g_c(\xi)}\,,\label{Tevals}
\eeq
and right eigenvectors corresponding to the non-zero eigenvalues 
\beq 
%e^{(0)}= \begin{pmatrix} 1\\0\\-1 \end{pmatrix},\; 
\mathbf e^{(1)}=\mathbf M= \frac{W_{0c}^2}{g_c}\begin{pmatrix} 1\\ \xi  \\ {g_c}W_{0c}^{-2}-1-\xi \end{pmatrix},\; \mathbf e^{(2)}= \frac{W_{0c}^2}{g_c}\begin{pmatrix} 1\\ \xi  \\ \lambda_2{g_c}W_{0c}^{-2}-1-\xi \end{pmatrix}.\label{Tevecs}
\eeq
Here 
$$
\mathbf M = \Omega^{-1}(\Omega_0,\Omega_1,\Omega_2)\,,
$$
where 
$$
\Omega = \sum_{i=0}^2 \Omega_i \,,
$$
such that $\sum_i M_i=1$.

There are now two cases to consider:
\begin{itemize}
\item[\underline{$\xi >\xi_c$}]

The eigenvalue $\lambda_2<1$, $\mathbb T_c$ is diagonalizable and it is straightforward to show that
 \bea\label{Hanswer}  {\mathbf H}^R=(1-\lambda_2)^{-1}
 \left[
 (1-(\lambda_2+1)g_cW_{0c}^{-1})\,{\mathbf M}
 + \lam_2^{R-1} (2g_cW_{0c}^{-1}-1)\,{\mathbf e}^{(2)}
\right].    \eea
 \item[\underline{$\xi =\xi_c$}]
 The eigenvalue $\lambda_2=1$ and we see from \eqref{Tevecs} that its eigenvector coincides with $\mathbf M$ so $\mathbb T_c$ has 
non-trivial Jordan normal form and a new vector 
 \beq \boldsymbol\varepsilon=\begin{pmatrix} 0\\0\\1\end{pmatrix}\eeq
 emerges satisfying
 \beq\label{Jshift} \mathbb T_c\, \boldsymbol\varepsilon=\mathbf M+\boldsymbol\varepsilon.\eeq
% $\mathbb T$ satisfies
%$$\mathbb T_c (\mathbb T_c-1)^2=0
%$$
%from which we find that, for $R\ge 1$,
%\beq\label{TR}
%\mathbb T_c^R = \mathbb T_c + (R-1)\mathbb T_c (\mathbb T_c-1) \,.
%\eeq
Setting 
$$
{\mathbf W} = (W_0,W_1,W_2)
$$
and noting that 
\beq\label{Wcreln}
{\mathbf W}_c=g_cW_{0c}^{-1}{\mathbf M}+(1-2g_cW_{0c}^{-1}){ \boldsymbol \varepsilon}
\eeq
then gives
\beq\label{HRc}
\mathbf H^R = \mathbf W_{c}  + (R-1)(1-2g_cW_{0c}^{-1})\mathbf M\,,\quad R\ge 1\,.
\eeq 
%where 
%\beq\label{Aeigc}
%a_0 = (1-\xi W_0^c) W_0^c\qquad a_2 = \xi W_0^2 + W_2^c\,.
%\eeq

 \end{itemize}
% 
% solve \eqref{Hrec} by diagonalizing $A$. This gives the following linear, 
%homogeneous recursion relation:
%\beq\label{HR} 
 %\begin{pmatrix} H_0^R \\ H_2^R\end{pmatrix} = \begin{pmatrix} e_0 \\ e_2\end{pmatrix} + \lam^{R-1} \begin{pmatrix} a_0 \\ 
%a_2\end{pmatrix}\,,
%\eeq
%where the eigenvectors $\begin{pmatrix} e_0 \\ e_2\end{pmatrix}$ and $\begin{pmatrix} a_0 \\ a_2\end{pmatrix}$ are determined by 
%the initial conditions \eqref{Hinit}. 
%$
%\begin{pmatrix} H_0^2 \\ H_2^2\end{pmatrix} = A \begin{pmatrix} W_0^c \\ W_2^c\end{pmatrix}\,.
%$
%One finds
%\beq\label{Aeig1}
%\begin{pmatrix} e_0 \\ e_2\end{pmatrix} = (1-\lam)^{-1} \begin{pmatrix} (1-\xi W_0^c)W_0^c \\ \xi W_0^2 + W_2^c\end{pmatrix}
%\eeq
%and 
%\beq\label{Aeig2}
%\begin{pmatrix} a_0 \\ a_2\end{pmatrix} = (1-\lam)^{-1} \begin{pmatrix} (\lam -\xi W_0^c)W_0^c \\ \xi W_0^2 + \lam W_2^c\end{pmatrix}
%\eeq

%\bigskip

\subsection{Volume of an infinite tree for $\xi_1=\xi_2=\xi$, $\xi_3=0$}

We let $K_i^R$ denote the (unnormalized) expectation value with respect to the measure $\mu_i$ of the number $k_R(\tau)$ of vertices at distance 
$R$ from the root of an infinite tree $\tau$ up to a normalization factor. Specifically,
$$
K_i^R = \Omega^{-1}\,\Omega_i\, \langle k_R \rangle_i\,.
$$
By decomposing the tree into its branches at the vertex $u_1$ next to the root one finds that $ K_i^R$ satisfies
\beq  K_i^{R}  =\frac{\partial F_i}{\partial W_j}  K_j^{R-1} + M_j\frac{\partial^2 F_i}{\partial W_j\partial W_k}  H_k^{R-1}\eeq
or
\beq {\mathbf K}^{R}  = \mathbb T_c {\mathbf K}^{R-1}+\Gamma {\mathbf  H}^{R-1}\label{Kprop}\eeq
where 
\beq \Gamma_{ik} = M_j\frac{\partial^2 F_i}{\partial W_j\partial W_k} = M_j\Lambda_{i,jk}. \eeq
The first term in \eqref{Kprop} is the contribution of the infinite branch and the second term that of the finite branches. As each tree has only a single vertex at height 1,
\beq {\mathbf K}^{1}={\mathbf M}.\eeq
The equation  \eqref{Kprop}  is easily iterated to get 
\bea {\mathbf K}^{R}  &=& {\mathbf M} + \sum_{\ell=1}^{R-1} \Trf_c^{\ell-1}\,\Gamma \,
{\mathbf  H}^{R-\ell}.
\label{Kformula}\eea
There are again two cases to consider:
\begin{itemize}
\item[\underline{$\xi >\xi_c$}] Combining \eqref{Hanswer} and \eqref{Kformula} and noting that 
\beq\label{GM}
\Gamma\, {\mathbf M}= 2g_c^{-1}W_{0c} \,{\mathbf M}\eeq
%$M$ is an eigenvector of $\Gamma$ with eigenvalue $2g^{-1}W_0$ *** is there a reason for this or is it a fluke? *** 
%
gives
\bea {\mathbf K}^{R}
&=& 2\,\frac{(W_{0c}g_c^{-1}-\lambda_2-1)}{1-\lambda_2} \,R\,\mathbf M + O(1)
\eea
from which  we get
\beq
\langle |B_R|\rangle_i = \Omega_i^{-1}\, \Omega\,\sum_{n=1}^R K_i^n = \Omega\,
\frac{(W_{0c}g_c^{-1}-\lambda_2-1)}{1-\lambda_2}\,R^2 +O(R).\label{Bresult}\eeq
It is straightforward to check that the coefficient of the $R^2$ term is positive for all  $\xi>\xi_c$ so we have shown 
that $d_h = 2$ in this regime. Note also that the coefficient diverges at  $\xi=\xi_c$, where $\lambda_2\to 1$, indicating 
that $d_h$ changes there.

 \item[\underline{$\xi =\xi_c$}] Combining \eqref{HRc} and \eqref{Kformula}  we have
 \begin{eqnarray}\label{basic} {\mathbf K}^{R} = {\mathbf M} &+& \sum_{\ell=1}^{R-1} \Trf_c^{\ell-1}\,\Gamma \left(
 (1-2g_cW_{0c}^{-1}) ({ \boldsymbol \varepsilon}-{\mathbf M}) +  g_cW_{0c}^{-1}{\mathbf M}\right)\nonumber\\
&+& \sum_{\ell=1}^{R-1} \Trf_c^{\ell-1}\,\Gamma \left( (R-\ell)(1-2g_cW_{0c}^{-1}){\mathbf M}  \right)
\end{eqnarray}
It is straightforward to check that at the tricritical point 
\beq 
\Gamma\, { \boldsymbol\varepsilon} = 2g_c^{-1}W_{0c}\left({\mathbf M} +\half { \boldsymbol\varepsilon}\right).
\eeq
Using this identity and \eqref{Jshift} then gives
\beq  {\mathbf K}^{R} =3 \left(\frac{W_{0c}}{2g_c}-1\right) R^2\, 
{\mathbf M} +O(R).\eeq
It is worth noting that one might have supposed from \eqref{Jshift}, \eqref{HRc} and \eqref {Kformula}  
%---------------NB NB: reference TR has been commented out!------------
% and \eqref{TR} 
%-----------------------
that ${\mathbf K}^{R}$ would be $O(R^3)$; 
however the coefficient of this leading term vanishes as a 
consequence of the tri-criticality condition. It follows now that
\beq
\langle |B_R|\rangle_i = \Omega_i^{-1}\, \Omega\, \sum_{n=1}^R K_i^n = \Omega \left(\frac{W_{0c}}{2g_c}-1\right) R^3\, +O(R^2)
\eeq
The coefficient of the $R^3$ term evaluates to a 
positive number so we have shown that $d_h = 3 $ at  $\xi=\xi_c$ in the AZ model.

\end{itemize}

\section{The extended model $\xi_3\ne0$}\label{sec:5}

%As remarked earlier the model with $\xi_1=\xi_2=\xi$ and $\xi_3=0$ that we have considered in detail so far is exactly the system that was discovered 
In  \cite{atkin-zohren} %(hereafter called the AZ model)
 it was argued that the AZ model with the CDT coupled to a reduced set of dimers is not likely to differ significantly from the full CDT-with-dimers system (hereafter called CDT-D).   We claim here that this is probably not correct by considering what happens when $\xi_3\ne0$; this perturbation is arguably closer to the CDT-D model as it incorporates more of the possible dimer types than the AZ model.

In the most general case $\mathbb T$ is given by
\begin{equation}\label{Tgen}
\mathbb T (\xi;g)= H 
\begin{pmatrix}
g(1+g\xi_3W_1)H & W_0 & W_0(1-g\xi_3W_0) \\
g\xi_1(1+g\xi_3W_1)H & W_1& W_1(1-g\xi_3W_0) \\
g\xi_2(1-W_1( 1-g\xi_3W_2 ))H & W_2     & \xi_2 W_0(1-W_1 -g\xi_3 W_0)\end{pmatrix}\;.
\end{equation}
It is straightforward although tedious to show that on the critical surface $g_c(\xi_1,\xi_2,\xi_3)$ 
\beq \mathbf M =  \frac{W_{0c}^2}{g_c}\left(1+\xi_3(1+\xi_1)W_0^3(2-g\xi_3W_0)\right)^{-1}\begin{pmatrix} 1\\ \xi_1  \\ {g_c}W_{0c}^{-2}-(1+\xi_1) (1-g\xi_3W_0)^2
\end{pmatrix}\eeq
 is always a right eigenvector with eigenvalue 1 and that the other eigenvalues are 0 (corresponding to the fact that $F_1=\xi_1 F_0$) and 
 \beq\label{lambda2gen} \lambda_2= \frac{\xi_1\xi_2 W_0^3}{g(1-g\xi_3W_0)^2} = g\xi_1\xi_2 W_0 H^2\,, \eeq
 where we have used 
$$
H = \frac{W_0}{g(1-g\xi_3 W_0)}\,.
$$
For definiteness we will first discuss the model with $\xi_1=\xi_2=\xi_3=\xi$ which, at least naively, is the closest we can get to CDT-D.  
For $\xi >\xi_{c} \approx -0.228$ we find that $g_c$ is a square root singularity of $W_0$, so $\alpha=\half$, and 
$\lambda_2<1$ at $g=g_c$.
Consequently  $m(g)\sim\vert\Delta g\vert^\half$ and  $d_H=2$  following the discussion of Section  \ref{sec:3}.  The infinite graph calculation of the local Hausdorff dimension follows the same lines as the $\xi>\xi_{c}$ case in Section \ref{sec:4} leading to $d_h=2$.  In this region of parameter space, where the dimer system is not critical, the model has exactly the same properties as the AZ model. 
 
However at  $\xi =\xi_{c}$ there is a tricritical point at which $\mathbb T_c$
 is diagonalisable,  $\lambda_2\approx 0.445  <1$ and $m(g)\sim \vert\Delta g\vert^\twothirds$ (the absence of a $\vert\Delta g\vert^\third$ term is  a consequence of the tricriticality condition).
Thus  $d_H=\threehalves$  but,  as shown in the Appendix, $d_h=1$.
% (the calculation again essentially  follows the  $\xi>\xi_{c}$ case in Section \ref{sec:4}).
 It is interesting to compare this result  with the simpler 
multicritical tree model of rooted binary trees with dimers 
placed on the edges, including the root edge, in such a way 
that no more than one dimer can end at any vertex 
\cite{aggs}. Letting $W_1$ and $W_0$ be the partition functions for trees with and without a dimer on the root edge respectively we see that they satisfy equations of the same form as \eqref{j3} but with
\bea F_0&=&g(W_0^2+2W_0W_1+1)\nn
F_1&=&g\xi(W_0^2+1).
\eea
This model also has a tricritical point with $\xi_c=-\frac{4}{27}$, exponent $\alpha=\frac{1}{3}$ and $\lam_{2c}<1$. One finds that   
\beq\label{j17}
 \Del\lam_1 \sim \frac{\prt g}{\prt W_0}\sim (\Del W_0)^2 \sim |\Del g|^{2/3}.
\eeq
which implies that  $d_H=3/2$ for $\xi=\xi_c$. On the other hand, using the results of the Appendix, $d_h=1$  as there is once 
again only one unit eigenvalue of $\mathbb T_c$.  The $\xi_1=\xi_2=\xi_3$ line of our model thus exhibits exactly the same behaviour 
as a standard multi-critical tree model.

 \begin{figure}[t]
\centerline{\scalebox{0.6}{\includegraphics{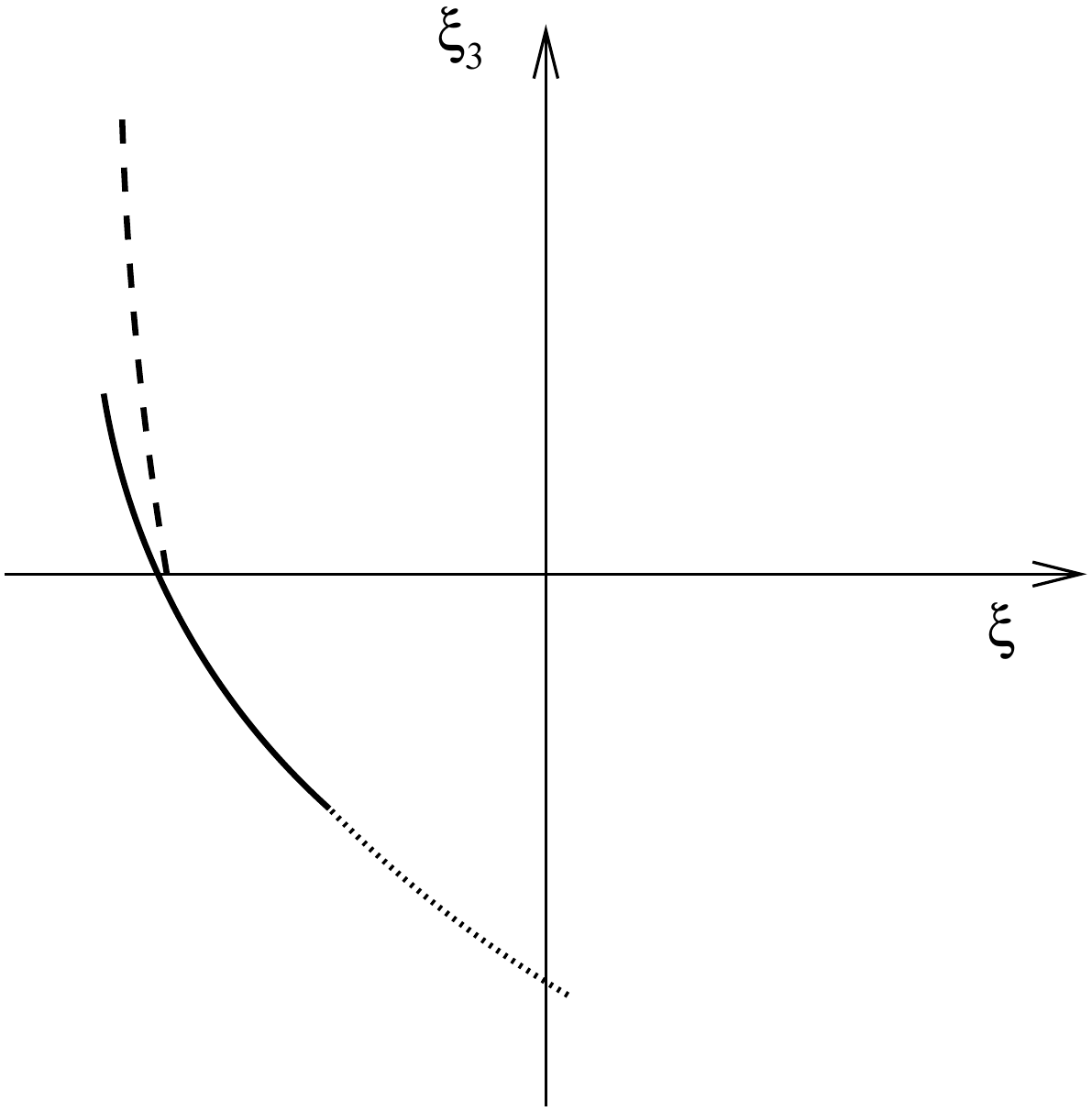}}}
\caption{The phase diagram in the $(\xi_1=\xi_2=\xi,\, \xi_3)$ plane. 
The solid line is the line of cubic degeneracies in $W_0$.
For $\xi_3 >0$ the physical region $S$ defined in Sec.\ \ref{sec:2}
lies to the right of the long dashed line, which is the line 
where $\lambda_2 =1$.  For $\xi_3 \leq 0$ the first part of the 
boundary of $S$ is the lower part of the solid line of 
cubic degeneracies in $W_0$. The dotted line makes up the rest of the lower 
part of the boundary of $S$. Along this part $W_0$ is only quadratic 
degenerate.}
\label{figPhaseDiagram}
\end{figure}

\begin{figure}
\centerline{\scalebox{0.6}{\includegraphics{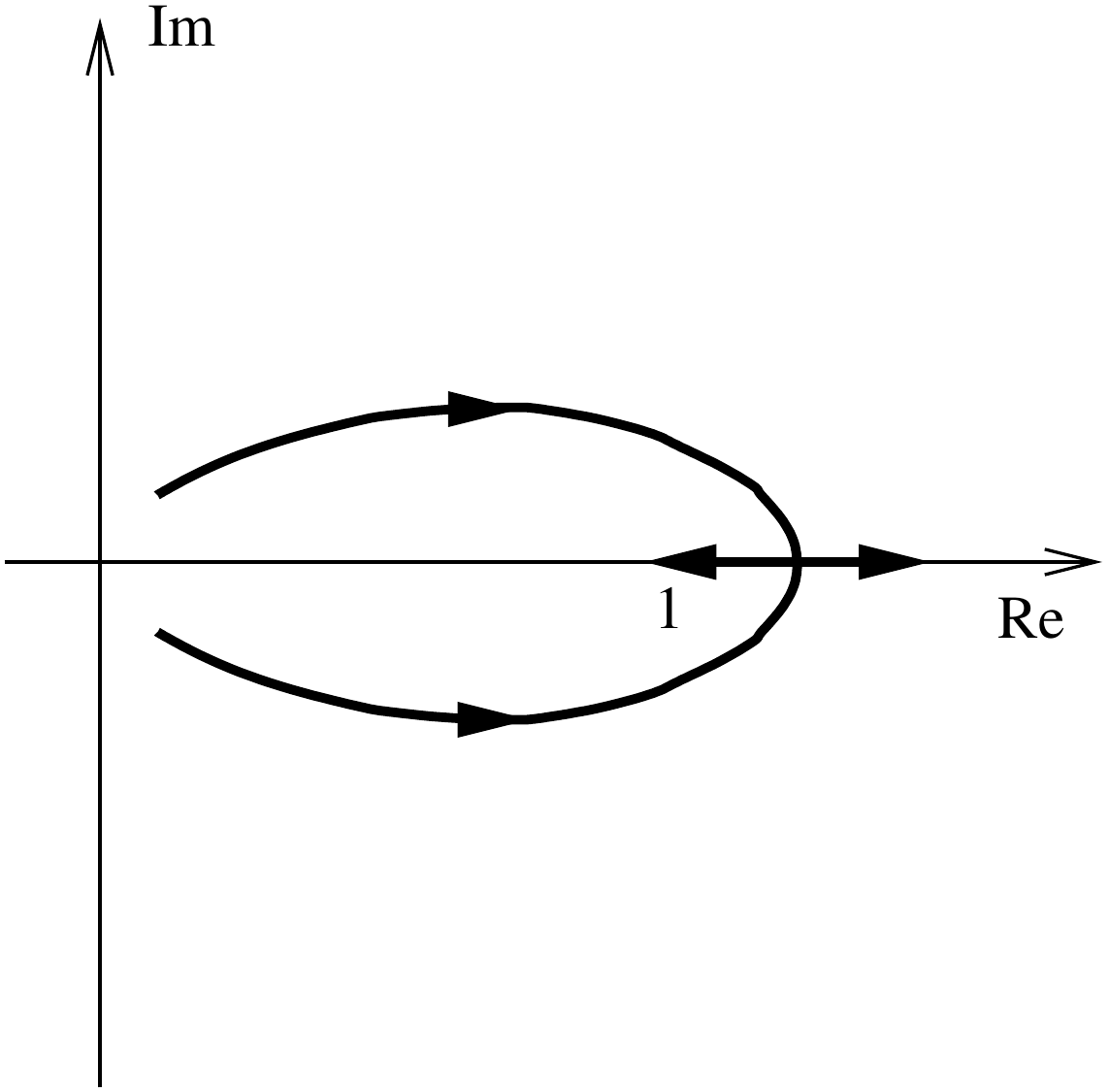}}}
\caption{The flow in the complex plane of the non-zero eigenvalues of $\bbT$; the arrow heads show how they move  as $g$ increases; $\lambda_1$ ends at 1, $\lambda_2$ at a value $>1$. }
\label{Tevalflow}
\end{figure}

 These calculations appear to show that the degenerate tri-critical point with $d_h=3$ found in the AZ model is very special and 
not at all characteristic of CDT dimer models in general. It is instructive to examine the phase diagram in the 
$(\xi_1=\xi_2=\xi, \,\xi_3)$ plane, see Fig \ref{figPhaseDiagram}.  There is a line of cubic degeneracies in $W_0$ that takes in 
the point $(-0.278\ldots, 0)$ and extends both above and below the $\xi$ axis. Using \eqref{lambda2gen} and the identity
$$
g^2\xi^2H^3 - g^2(1+\xi)\xi_3 H^3 -1=0
$$
which holds at tricritical points as a consequence of \eqref{j9},\eqref{j10} and \eqref{j11}
it follows that
 $$
\lambda_2 = 1 + g_c \xi_3 (1+\xi - \xi^2 W_0) H^3\,.
$$
Since the expression in parenthesis, $H$ and $g_c$ are all positive it follows that on the tricritical line $\lambda_2 <1$ 
for $\xi_3<0$ but that  $\lambda_2 >1$ for $\xi_3>0$. The latter behaviour is a little strange; it would in fact be a contradiction 
for  a purely real  eigenvalue of $\bbT$ to go through 1 \emph{before} criticality is reached
 (see the Appendix for example).   Closer inspection shows that at small $g\ll g_c$ the eigenvalues are complex and as $g$ 
increases they flow as shown in Fig \ref{Tevalflow}.  However exponential growth of the two-point function is a symptom that the 
series in $\xi_i$ and $g<g_c$ for $Z$  is not absolutely convergent and the effect of the negative weights is 
sufficiently strong that the conventional statistical 
 mechanical interpretation of the model fails. We conclude that the physical region for $\xi_3>0$ extends only as far as the 
line where $\lambda_2 =1$ at $g_c$. It
 can be checked that inside the region and along this line there are only quadratic degeneracies in $W_0$ and that $\mathbb T_c$ 
is diagonalisable so $d_h=d_H=2$. On the other hand for $\xi_3<0$ there is a genuine line of tricriticality which includes 
the $\xi=\xi_3$ point analysed above and ends at $(-0.162\ldots,-0.582\ldots)$. Beyond this point the tricriticality 
disappears and even on the boundary of the physical region $\alpha=\half$ and $d_h=d_H=2$.

% \begin{figure}[t]
%\centerline{\scalebox{0.6}{\includegraphics{PhaseDiagram.pdf}}}
%\caption{The phase diagram in the $(\xi_1=\xi_2=\xi,\, \xi_3)$ plane. The solid line is the line of cubic degeneracies in $W_0$; the long dashed line is the line where $\lambda_2 =1$; and the dotted line the boundary of the physical region. }
%\label{figPhaseDiagram}
%\end{figure}
 
 %?? should probably add a disclaimer about what we have not investigated in detail ??

%perturbation to $\eta=\xi\ne0$ closer to real dimer model

%phase diagram

%physical region

%behaviour of eigenvalues

%uniqueness of tri-crit

\section{Concluding remarks}

At the critical value of the dimer fugacity we expect that CDT-D, the full dimer model on CDT,  represents a lattice regularization of projectable Ho\v{r}ava-Lifshitz quantum gravity 
coupled to a (2,5) minimal conformal field theory. 
In the absence of a solution to  CDT-D we have obtained 
the solution of a restricted dimer model and mapped out its phase diagram. In particular, 
we have seen that the geometric features of the AZ model  \cite{atkin-zohren} are very 
special and not robust under perturbations. At generic points of the phase 
boundary we have found the values of the Hausdorff dimensions are either $d_H=3/2$ and
 $d_h=1$, coinciding with the values for the simplest multicritical tree 
\cite{adj,book},  or $d_H=d_h=2$.  

One may speculate on the implications of our results for the CDT-D model. 
While we do not have any rigorous results in this direction it is 
worth noting that 
the full dimer model on a generalized CDT \cite{GCDT} has been solved 
exactly in \cite{aggs} using a matrix model representation yielding 
the value $d_H=3/2$. 
The generalized causal triangulations of this model 
can be defined combinatorially 
\cite{ab} or by using a special scaling limit of matrix models \cite{GCDT}.
This slightly more general set of triangulations has many of the 
characteristics of CDT, e.g.\ $d_H=d_h=2$. Hence it is tempting on 
the basis of this result 
to conjecture that $d_H=3/2$ and $d_h=1$ are indeed the correct 
values for the full dimer model on a CDT.

It is natural to extend the  above considerations further. One can 
define multicritical generalized CDT models \cite{az-multi}, which 
most likely correspond to specific fine-tuned scaling limits of 
matrix models, generalizing the considerations in \cite{aggs}.
Recall that the standard multicritical matrix models from DT provide 
representations of 2d Euclidean quantum gravity coupled to 
certain conformal field theories. They also have the interpretation
of (increasingly complicated) fine-tuned multi-dimers systems on the
DT-set of random graphs. Thus it is possible that the multicritical 
behavior found in \cite{az-multi} represents the effect of 
fine-tuned multi-dimer models on the generalized CDT-set of random 
graphs, and has the continuum interpretation of certain conformal
field theories coupled to  2d Ho\v{r}ava-Lifshitz gravity.

This leave us with the interpretation of the AZ model. Although this 
model is special, it is not {\it that} special. As we have seen there 
is a least a one-parameter set of coupling constants leading to the same 
scaling. Thus we believe there should also be a 
continuum interpretation of this class of models. 

%\vspace{.5cm}      
\subsection* {\ Acknowledgements.} 
J.A.  acknowledges support from the ERC Advanced Grant 291092 
``Exploring the Quantum Universe'' (EQU) as well as from
the Free Danish Research Council  grant ``Quantum gravity and the
role of black holes.'' In addition JA was supported 
in part by the Perimeter Institute of Theoretical Physics.
Research at Perimeter Institute is supported by the Government of Canada
through Industry Canada and by the Province of Ontario through the 
Ministry of Economic Development \& Innovation. 
J.A. and B.D. acknowledge support from NordForsk researcher 
network Random Geometry (grant no. 33000)".

%\appendix
\section*{Appendix: \boldmath{$d_h$} and the multi-critical condition }

We discuss here how the condition for multi-criticality affects $d_h$ in general. We consider a set of generalised trees with first vertex label $\ell(v_0)=i$ whose partition functions
$W_i$ satisfy
\beq W_i = F_i(\mathbf W;\vxi;g),\qquad i=1\ldots N.\eeq
%
%where $\xi$ is some dimer fugacity.
%
The approach to criticality in the grand canonical ensemble (GCE) is governed by (repeated indices are summed over)
\beq \delta W_i = \delta g \frac{\partial F_i}{\partial g}+ \bbT_{ij}\delta W_j + \half \Lambda_{i,jk}\,\delta W_j \delta W_k+{\mathrm {h.o.t.}}\eeq
where
\bea  \bbT_{ij}= \frac{\partial F_i}{\partial W_j} ,\quad
\Lambda_{i,jk}=\frac{\partial^2F_i}{\partial W_j\partial W_k},\quad
(\mathbf\Lambda_{jk})_i=\Lambda_{i,jk}.
\eea
Re-arranging
\beq ((1-\bbT)\delta W)_i = W_i\delta g  + \half \Lambda_{i,jk}\,\delta W_j \delta W_k+{\mathrm {h.o.t.}}\label{GCE}\eeq
so criticality is reached as $g\uparrow g_c$ where the largest real eigenvalue of $\bbT$ first reaches 1.

Now, setting $g=g_c$ and working in the critical ensemble, consider the infinite single spine trees with first vertex labelled by $\ell(u_1)=i$. Decomposing these trees at their first vertex $u_1$ into an infinite component and finite components we see that their measures $M_i$ satisfy 
\beq\label{M_evector} M_i=\left(\frac{\partial F_i}{\partial W_j}\right)_c M_j = \bbT_{cij} M_j.\eeq
We will assume that they are normalised so that the total measure is 1,
\beq 1=\sum_i M_i.\eeq
%
%i.e. the total measure is 1. Then, setting $g=1$ and working in the critical ensemble, $M_i$ satisfies 
%
%\beq M_i=\frac{\partial f_i}{\partial W_j} M_j\eeq
%
%i.e. $M_i$ is the unit eigenvector of the matrix
We see that $\bbT$ directly relates the infinite spine trees and the GCE. At criticality it must have at least one eigenvalue which is one and this must be the first real eigenvalue to reach one, otherwise the system would have reached criticality at some smaller value of $g$.

Turning to the local Hausdorff dimension we note from \eqref{Kformula}, \eqref{H3} and  \eqref{M_evector} that
%Let $ H_i^R$ denote the expectation value of the number of vertices at height $R$  on a finite tree at criticality; then by decomposing the tree at its first vertex $v_0$ we find that
%
%\bea \mathbf H^{R}=\bbT_{c}\mathbf H^{R-1}
%				= \bbT_{c}^{R-1}\mathbf W_c  \eea
%
%so
%
%\bea H_i^{R}&=&(\bbT^R)_{ij}H_j^{0}  \\
%&=& (\bbT^R)_{ij}W_j \eea
%
%on account of the fact that each tree has exactly one vertex at height 1. Now let 
$K_i^R$, the  expectation value with respect to the measure $\mu_i$ of the number of vertices at distance $R$ from the root of an infinite tree,  %(here unnormalized means that the factor $\Omega_i^{-1}$ in \eqref{mui} is omitted);
 %then again by decomposing the tree at its first vertex $u_1$ we find that
%
%Now let $ K_i^R$ denote the number of edges  at height $R$ ($R=0$ being the root edge)  on an infinite spine tree. It satisfies
%
%\beq  K_i^{R+1}  =\frac{\partial f_i}{\partial W_j}  K_j^{R} + M_j\frac{\partial}{\partial W_k}\left(\frac{\partial f_i}{\partial W_j}\right)  H_k^{R}\eeq
%
%or
%
is given by
%\beq {\mathbf K}^{R+1}  = \bbT {\mathbf K}^{R}+\Gamma {\mathbf  H}^{R}, \label{KpropA}\eeq
%
%
%The first term in \eqref{Kprop} is the contribution of the infinite branch and the second term of the finite branches. 
%Again as each tree has only a single vertex at height 1, ${\mathbf K}^{0}={\mathbf M}$,
%
%\beq {\mathbf K}^{0}={\mathbf M}.\eeq
%
%and  from \eqref{KpropA} and \eqref{M_evector}  we get 
%
%\bea {\mathbf K}^{R}  &=& \bbT^R \,{\mathbf M} + \sum_{\ell=1}^R \bbT^{\ell-1}\,\Gamma \,\bbT^{R-\ell}\,{\mathbf W}.\nonumber \\
%
\bea {\mathbf K}^{R} &=&{\mathbf M} + \sum_{\ell=1}^{R-1} \bbT^{\ell-1}\,\Gamma \,\bbT^{R-\ell-1}\,{\mathbf W},
\label{Kanswer}\eea
where 
\beq \Gamma_{ik} = M_j\frac{\partial}{\partial W_k}\left(\frac{\partial F_i}{\partial W_j}\right)= M_j\Lambda_{i,jk}. \eeq

%\section{Hausdorff Dimension}

%Before proceeding let's check this against the usual Galton-Watson tree with generating function $f(x)$ and $f'(1)=f(1)=1$. Then at criticality
%
%\beq W=f(W)=1\eeq
%
%and
%
%\beq \bbT_{11}=\pdiff{f}{W}=1\eeq
%
%so $M=1$ is consistent and
%
%\beq H^R=1\eeq
%
%Then
%
%\beq \Gamma_{11}=M\pdifftwo{f}{W}=f''(1)\eeq
%
%so
% 
%\beq K^R=1+R\,f''(1)\eeq
%
%which is the correct result. Note that any single component multi-critical tree has $f''(W_{crit})=0$ as the first multi-criticality condition and therefore $d_H=1$.

%We'll do some detailed calculations in the following sub-sections but clearly 

The implications in general of \eqref{Kanswer} for the Hausdorff dimension depend very much upon the Jordan decomposition  of  $\bbT = S J S^{-1}$ where $J$ is of Jordan Block form $\mathrm{Diag}(J^{<},J^1)$; the block $J^1$ corresponds to $r$ unit eigenvalues and  $J^{<}$ to the $N-r$ eigenvalues which are less than 1. If $J^1$ is diagonal then
\beq\label{Jpower} J^\ell = \mathrm{Diag}(0,\ldots 0,1,\ldots 1) +O(\alpha^\ell)\eeq
where $\alpha$ is the largest eigenvalue smaller than 1. 

%In this case ${\mathbf K}^{R}$ varies at most linearly with $R$ -- although as we'll show soon the coefficient may vanish.  Only if $J^1$ is a non-trivial Jordan block does $\bbT^R$ grow with $R$ at a rate which depends on the size of the non-trivial block -- linear for the smallest, quadratic for the next largest and so on.

%\subsection{Single unit eigenvalue}
We first consider the case where there is a single unit eigenvalue so $J^1=1$. Introduce the orthonormal basis $(\e^i)_j=\delta_{ij}$ so that
\bea \delta \W &=& S (\mu\, \e^N+ \nu^a \,\e^a)\,, \qquad
 \W = S (A\, \e^N+ B^a \,\e^a)\eea
where $a, b =1\ldots N-1$. $A\ne 0$ and $B^a$ are constants and up to normalisation the measure vector is
\beq \M=S\,\e^N.\eeq
Substituting in \eqref{GCE} we obtain
%
%\bea S(1-J)\nu^a \,\e^a &=& S (A\, \e^N+ B^a \,\e^a)\delta g  + 
%\half \mathbf\Lambda_{jk}\,   (\mu\, \M+ \nu^a S\,\e^a)_j    (\mu\, \M+ \nu^b S\,\e^b)_k +{\mathrm {h.o.t.}}\nonumber\\ 
%&=&S (A\, \e^N+ B^a \,\e^a)\delta g+\half\mu^2\Gamma\M+\mu\nu^a\Gamma S\,\e^a
%+\half\nu^a\nu^b\mathbf\Lambda_{jk} (S\,\e^a)_j (S\,\e^b)_k \nonumber\\
%\eea
%
%or
%
\bea (1-J^<)\nu^a \,\e^a &=& (A\, \e^N+ B^a \,\e^a)\delta g+ \half\mu^2S^{-1}\Gamma S \e^N+\mu\nu^aS^{-1}\Gamma S\,\e^a\nonumber\\
&&+\half\nu^a\nu^b S^{-1}  \mathbf\Lambda_{jk} (S\,\e^a)_j (S\,\e^b)_k\,.\label{GCEmcrit}
\eea
Since $1-J^<$ is invertible we see that, provided $ (S^{-1}\Gamma S )_{NN}$ is non-zero, 
$\mu\sim (\delta g)^\half$ and criticality is quadratic. The multi-critical condition is 
\beq (S^{-1}\Gamma S )_{NN}=0.\label{multicrit}\eeq
Now returning to \eqref{Kanswer}
\bea {\mathbf K}^{R}  
&=&{\mathbf M} + \sum_{\ell=1}^{R-1} S J^{\ell-1}\,S^{-1}\Gamma \,SJ^{R-\ell-1}S^{-1}\,{\mathbf W}
\eea
%and
%
%\beq (J^k)_{ij}=\delta_{iN}\delta_{jN} +O(\alpha^k)\eeq
whence, using \eqref{Jpower},
\bea {\mathbf K}^{R}  
&=&{\mathbf M} + R \,\boldS \,(S^{-1}\Gamma \,S)_{NN} (S^{-1}\,{\mathbf W})_N +O(1)
\label{Kresult}\eea
 where $ \boldS_i=S_{iN}$.
We see from \eqref{multicrit} and \eqref{Kresult} that the linear term linear in $R$ automatically vanishes at the multi-critical point where, therefore, $d_h=1$.  This is completely standard multi-criticality and the result is identical to that for the single component multi-critical tree. It is straightforward  to generalise this analysis to models where $J^1$ is of higher rank but still diagonal and find the same conclusion that $d_h=1$.  Note that it is always necessary to compute the actual coefficient of the remaining leading term in a particular model to check that it is positive otherwise the result is meaningless.

%Jan's Polymer Dimer Model is of this type. We have
%
%\bea W_1&=&g(W_1^2+2W_1W_2+1)\nn\\
%W_2&=&g\xi(W_1^2+1)\eea
%
%and
%\bea \bbT=2g\left(\begin{array}{cc} W_1+W_2&W_1\\ \xi W_1&0\end{array}\right).\eea
%
%It is obvious that $\bbT$ has a normal Jordan form because it can be turned into a symmetric matrix by the similarity transformation
%
%\beq \left(\begin{array}{cc} 1&0\\  0&\xi^\half\end{array}\right).\eeq

%\subsection{Multiple unit eigenvalue, non-diagonal Jordan form}

The situation is different  if $J^1$ is non-diagonal.  We will analyse the simplest case where there are two unit eigenvalues and we have the simplest non-diagonal Jordan block
\bea J^1=\left(\begin{array}{cc} 1&1\\ 0&1\end{array}\right).\eea
%
%Actually, the way I've set things up,  it would be marginally more convenient to have the block lower diagonal but probably one should stick to the standard notation.
%Anyway  
We then have that 
%
%\beq J^\ell =  \left(\begin{array}{ccc} \ddots&0&0\\  0&1&\ell\\0&0&1\end{array}\right) +O(\alpha^\ell)\eeq
%
%where the dots denote the subspace with max eigenvalue $\alpha <1$. So
%
\beq\label{JBform} (J^\ell )_{ij}=\delta_{i,N-1}\delta_{j,N-1}+\delta_{i,N}\delta_{j,N}+\ell\delta_{i,N-1}\delta_{j,N} + O(\alpha^\ell).\eeq
$\M$ is the ordinary eigenvector with eigenvalue 1,
\beq \M=S\e^{N-1}\eeq
but now there is a vector belonging to the second eigenvalue 1
\beq \boldsymbol \varepsilon=S\e^N\eeq
with the property
\beq \bbT\, \boldsymbol \varepsilon=\M+\boldsymbol \varepsilon.\eeq
Now we have
\bea \delta \W &=& S (\mu\, \e^{N-1}+\lambda\,\e^{N}+ \nu^a \,\e^a)\, , \quad
 \W = S (A\, \e^N+C\, \e^{N-1}+B^a \,\e^a)\,,\eea
where $a, b =1\ldots N-2$. $A\ne 0$ and $B^a$ are constants
and substituting in \eqref{GCE} we find
\bea (1-J^<)\nu^a \,\e^a+\lambda \,  \e^{N-1} &=&  (A\, \e^N+C\, \e^{N-1}+ B^a \,\e^a)\delta g  + \nn
&& +\half S^{-1}\mathbf\Lambda_{jk}\,   (\mu\, \M+ S \lambda\,\e^{N}+    \nu^a S\,\e^a)_j \times \, \nn &&S (\mu\, \e^{N-1}+\lambda\,\e^{N}\mu\, \M+ \nu^b \e^b)_k +{\mathrm {h.o.t.}}\label{GCEmcritJworking}
\eea
and closing with $\e^N$,
\bea 0&=& A\delta g+ \half\mu^2(S^{-1}\Gamma S)_{N N-1}+\ldots \label{GCEmcritJ}
\eea
showing that 
\bea (S^{-1}\Gamma S)_{N N-1}=0\eea
is a necessary condition for multi-criticality. Note that because of the Jordan block structure $\lambda$  appears linearly on the l.h.s.  of \eqref{GCEmcritJworking} so the leading singularity can only be associated with $\M$, and not with $\boldsymbol\varepsilon$. 
Using \eqref{Kanswer}  and \eqref{JBform} we get 
\bea {\mathbf K}^{R} _i
&=&{\mathbf M}_i + \sum_{\ell=1}^{R-1}  S _{ij'}
(   \delta_{j',N-1}\delta_{j,N-1}+\delta_{j',N}\delta_{j,N}+(\ell-1)\delta_{j',N-1}\delta_{j,N}         )\nn
&&\,(S^{-1}\Gamma \,S)_{jn}
(\delta_{n,N-1}\delta_{m,N-1}+\delta_{n,N}\delta_{m,N}+(R-\ell-1)\delta_{n,N-1}\delta_{m,N} )
(S^{-1}\,{\mathbf W})_m+O(1)\nn
&=&{\mathbf M}_i + \sixth (R-1)(R-2)(R-3)    S_{iN-1} (S^{-1}\Gamma \,S)_{N N-1} (S^{-1}\,{\mathbf W})_N\nn
&&+\half (R-1)(R-2) \sum_{L=N-1}^N \left[ S _{iN-1} (S^{-1}\Gamma \,S)_{N L}(S^{-1} \,{\mathbf W})_{L} +
\,S_{iL}(S^{-1}\Gamma \,S)_{L N-1}(S^{-1}\,{\mathbf W})_{N}\right]\nn
&&+O(R)\label{KmcritJ}
\eea
We see from \eqref{KmcritJ} that the multi-critical condition automatically suppresses the $(R-1)(R-2)(R-3)$ in $\K^R$ but that the quadratic $(R-1)(R-2)$ term survives.  Hence $d_h=3$, again provided that the numerical coefficient, which has to be  computed in a particular model, is positive.

\end{document}